\documentclass[11pt, a4paper]{article}
\usepackage{jheppub}
\usepackage{subcaption}
\usepackage{multirow}
\usepackage{multicol}
\usepackage[normalem]{ulem}
\usepackage{float}
\usepackage{caption}
\usepackage[english]{babel}
\usepackage{adjustbox}
\usepackage{tikz-feynman}
\usepackage{contour}
\usepackage[toc,page]{appendix}
\usepackage{url}
\usepackage{slashed}

\newcommand{\br}{\begin{eqnarray}}
\newcommand{\er}{\end{eqnarray}}

\usepackage{tikz,xcolor,hyperref}

\definecolor{lime}{HTML}{A6CE39}
\DeclareRobustCommand{\orcidicon}{\hspace{-4pt}
\begin{tikzpicture}
	\draw[lime, fill=lime] (0,0) 
	circle [radius=0.16] 
	node[white] {\hspace{0.1mm}{\fontfamily{qag}\selectfont \tiny ID}};
	\draw[white, fill=white] (-0.07,0.1) 
	circle [radius=0.01];
\end{tikzpicture}
\hspace{-3.2mm}
}

\foreach \x in {A, ..., Z}{\expandafter\xdef\csname 
orcid\x\endcsname{\noexpand\href{https://orcid.org/\csname orcidauthor\x\endcsname}
	{\noexpand\orcidicon}}
}

\newcommand{\met}{\slashed{E}_T}

\title{Probing axion-like particles through associated gauge-boson recoil and polarization at the LHC}
	\author[a]{\small Amandip De\orcidA{},}
\emailAdd{deamandip93@gmail.com}
\affiliation[a]{Department of Physics, SRM University-AP, Amaravati 522240, Andhra Pradesh, India}

\author[b]{\small Monoranjan Guchait\orcidB{},}
\emailAdd{guchait@tifr.res.in}
\affiliation[b]{Tata Institute of Fundamental Research, Homi Bhabha Road, Colaba, Mumbai 400005, India}

\author[c,d]{\small and Arnab Roy\orcidC{},}
\emailAdd{arnab.roy1@monash.edu}
\affiliation[c]{School of Physics and Astronomy, Monash University,
Wellington Road, Clayton, Victoria 3800, Australia}
\affiliation[d]{Departament de F\'isica Qu\'antica i Astrof\'isica, Institut de Ci\`encies del Cosmos (ICCUB), Universitat de Barcelona, Mart\'i Franqu\`es 1, E08028 Barcelona}

\abstract{
Axion-like particles (ALPs) appear in many well-motivated extensions of the Standard Model (SM) and can exhibit characteristic interactions with electroweak gauge bosons ($V = W^\pm/ Z$) through higher-dimensional operators. We explore the production of ALPs ($a$) in association with $V$ at the HL-LHC, with $\sqrt{s}=14~{\rm TeV}$ and an integrated luminosity of $\mathcal{L}=3~{\rm ab}^{-1}$. We study the process $pp\to Va$, with $a\to\gamma\gamma$, for ALP masses in the range $20~{\rm GeV}\leq m_a\leq 200~{\rm GeV}$. This analysis extends existing associated-production searches by exploiting characteristic recoil and polarization features induced by the momentum-dependent $\rm {V-V-a}$ interaction. Using both cut-based selections and an \texttt{XGBoost} multivariate analysis, we demonstrate that these observables provide robust discrimination against SM backgrounds beyond the diphoton mass peak, improving the sensitivity to electroweak ALP couplings to $C_{WW} \gtrsim 5.6 \times 10^{-3}~\mathrm{TeV}^{-1}$ and $C_{BB} \gtrsim 2.8\times 10^{-2}~\mathrm{TeV}^{-1}$. We further show that, if a resonance peak is observed in data, the same observables can help distinguish an ALP-like signal from a Higgs-like spin--0 state of the same mass.}

\begin{document}
\maketitle
\flushbottom

\section{Introduction}
\label{sec:intro}
Axion-like particles (ALPs) have received a lot of attention in recent years as one of the well-motivated candidates for physics beyond the Standard Model (BSM). The theoretical motivation arises naturally from several extensions of the Standard Model (SM); e.g., see
Refs~\cite{DiLuzio:2020wdo,Choi:2020rgn} for a detailed review.
Historically, the axion was originally introduced as a solution to the strong CP problem~\cite{Peccei:1977hh,Peccei:1977ur,Weinberg:1977ma,Wilczek:1977pj}. The Lagrangian is required to be symmetric under global 
Peccei--Quinn (PQ) symmetry, namely $U(1)_{PQ}$, which is broken at some high energy scale, and a Nambu-Goldstone boson (pNGB) appears as a result of this broken symmetry, referred to as QCD axions.
Non-perturbative QCD relates the axion mass to the
PQ scale, $m_a\simeq 5.7~\mu{\rm eV}(10^{12}{\;\rm GeV}/f_a)$, so the conventional
invisible axion is very light and weakly coupled
\cite{Kim:1979if,GrillidiCortona:2015jxo,DiLuzio:2020wdo,Choi:2020rgn}. 
Consequently, only the decays to photons, neutrinos, and gluons are accessible. 
From a cosmological perspective, the decay of axions or ALPs into photons is particularly interesting. A variety of cosmological and astrophysical studies have placed stringent constraints on the axion--photon coupling strength; see, for example, Ref~\cite{AxionLimits}.

In general, ALPs can arise as pNGBs from the spontaneous breaking of approximate global symmetries and need not be directly related to the strong CP problem~\cite{DiLuzio:2020wdo,Choi:2020rgn}. Unlike the QCD axion, their
masses and couplings are not fixed by the QCD axion mass-coupling relation, allowing a broad range of phenomenologically viable parameter space. This makes ALPs 
particularly interesting candidates for new physics and motivates their study in a variety of theoretical and experimental contexts.

One of the most appealing aspects of axions and ALPs is their potential to provide a viable dark matter candidate, with profound implications for cosmology ~\cite{Preskill:1982cy, Abbott:1982af, Dine:1982ah, Chadha-Day:2021szb}. Moreover, ALPs can arise in dark-matter portal scenarios, where they act as mediators between the dark sector and SM  particles~\cite{Nomura:2008ru,Fitzpatrick:2023xks}.

Since the mass and couplings of ALPs are free parameters, they can be 
looked for over a broad range of masses, from the MeV scale to the electroweak scale and beyond through their couplings to photons, gluons,
fermions and electroweak gauge bosons. Therefore, ALPs can be accessible in collider experiments. At the LHC, light pseudoscalars equivalent to 
ALPs have been searched for in a variety of photon-rich final states. Resonant diphoton searches constrain prompt $a\to\gamma\gamma$ decays, including scenarios where ALP is in boosted low-mass regime
\cite{ATLAS:2022abz}. Searches for exotic Higgs decays probe
$H\to aa\to4\gamma$ and $H\to Za$ with $a\to\gamma\gamma$
\cite{ATLAS:2023ian,CMS:2022xxa,ATLAS:2023etl}.  
Pair production of light bosons in the exotic decays of the Higgs boson results in a wide range of final states in various combinations of fermions and photons. For instance, ATLAS and CMS have carried out extensive searches for light bosons
in the $bb\tau\tau$, $b b \mu\mu$,  $\mu\mu\tau\tau$ $\gamma\gamma  jj$, $4b$, $4\gamma$ and $b\bar b \gamma\gamma$ and 
$\gamma\gamma j j $ final states~\cite{CMS:2018zvv,ATLAS:2015unc,CMS:2018nsh,ATLAS:2021hbr,CMS:2024zfv, ATLAS:2020ahi,CMS:2022xxa,CMS:2022fyt,ATLAS:2024nnm,ATLAS:2018jnf,CMS:2026elu}. Complementary limits arise from LEP and LHC
studies of $Z\to a\gamma$ or multi-photon final states
\cite{Jaeckel:2015jla}.
Phenomenologically, several studies have explored the prospects for ALP searches at the LHC in association with photons, jets, gauge bosons, and top quarks and have derived constraints on the ALP mass and its couplings using LHC data 
~\cite{Mimasu:2014nea,Brivio:2017ije,Bauer:2017ris,Wang:2021uyb,Gavela:2019cmq,Carra:2021ycg,Ren:2021prq,AlonsoAlvarez:2018irt,Cheung:2024qge,Anuar:2024qsz}.
ALP production through photon fusion has also been constrained using light-by-light scattering in ultraperipheral heavy-ion collisions
\cite{ATLAS:2020hii,CMS:2018erd}.  
In the absence of any observed ALP signal, exclusion limits on the couplings of ALPs to the respective SM particles are generally presented as a function of the ALP mass($m_a$).
For example, current experimental bounds are $g_{a\gamma\gamma} \sim 4 \times 10^{-5}~ \mathrm{GeV}^{-1}$ \cite{ATLAS:2023zfc} for 
($m_a$=150-1600) GeV, $g_{aZ\gamma} \sim 4 \times 10^{-4}~ \mathrm{GeV}^{-1}$ \cite{Carra:2021ycg}, $g_{aZZ} \sim 6.6 \times 10^{-4}~ \mathrm{GeV}^{-1}$ \cite{CMS:2021xor}, $g_{aWW} \sim 5 \times 10^{-4}~ \mathrm{GeV}^{-1}$ \cite{Carra:2021ycg}. 
Potentially, ALPs can be very light, even below the QCD scale, and are therefore subject to strong constraints from low-energy observables and cosmological measurements. For lighter ALPs, collider searches are complemented by stringent constraints from beam-dump experiments, flavor observables, stellar cooling, supernovae, and cosmological observations. These probes exclude substantial regions of the ALP parameter space, particularly in the $(m_a,g_{a\gamma\gamma})$ and $(m_a,g_{ae})$ planes~\cite{Jaeckel:2010ni,Irastorza:2018dyq,Agrawal:2021dbo,AxionLimits}.

ALPs have also been widely searched for at the LHC in associated-production channels such as $pp\to Za$ and $pp\to W^\pm a$, typically using rates and standard kinematic distributions in specific decay modes such as $a\to\gamma\gamma$
\cite{Mimasu:2014nea,Bauer:2018uxu,AlonsoAlvarez:2018irt,Brivio:2017ije, Bauer:2017ris,Wang:2021uyb,Cheung:2024qge,Chenarani:2025uay,
        Gavela:2019cmq,Bonilla:2022pxu}. 
        In this current study, we focus on the ALP 
production in association with a gauge boson
(V=W,Z) in $pp$ collisions at the high luminosity LHC (HL-LHC) 
with the center of energy 14 TeV.
Most studies~\cite{Cheung:2024qge,Gavela:2019cmq} of associated production exploit the decay products of 
accompanying $W$ or $Z$ boson primarily as a tag, mainly 
leptons as a clean trigger, and also to suppress backgrounds. 
We follow a complementary approach focusing on leptons as a 
polarimeter to exploit the polarization of the associated gauge boson.
The tensor structure of the ALP couplings to associated gauge bosons leads to distinct polarization and kinematic properties. Schematically, a Higgs-like interaction $H V_\mu V^\mu$, and 
an ALP-like interaction $a V_{\mu\nu}\widetilde V^{\mu\nu}$
populate the polarization states and transverse recoil 
of the associated gauge boson in characteristically different ways.
The polarization of the associated $W$ or $Z$ can be manifested
from the charged-lepton angular distributions in the gauge-boson rest frame, allowing one to construct polarization fractions, 
angular coefficients, and asymmetries. 
Such a kind of polarimetry has been developed in detail, e.g 
for the SM gauge boson production to 
probe its V-A structure of couplings, and   
Higgs studies for constraining non-standard $HVV$ tensor structures 
\cite{Nakamura:2017ihk,Goncalves:2018fvn,Rao:2018abz,Colyer:2025ehv,
	Rao:2020hel,Ellis:2013ywa,Godbole:2013saa,Godbole:2014cfa,
	Maltoni:2013sma,CMS:2016tad,CMS:2019jdw}. 
Here we use the same principle of the polarimetric idea in a 
different context: not as a standalone discovery observable, 
but as part of a broader set of physically motivated variables 
for understanding and discriminating spin--0 resonance.

This motivates two classes of observables tied directly to the 
production vertex. The first consists of polarimetric variables, 
such as the charged-lepton helicity angle $\theta^\ast$ in the $V$ rest 
frame (or equivalently, di-photon rest frame), 
which probes the longitudinal and transverse polarization 
components of the associated gauge boson. The second consists of 
recoil observables, $p_T^V$, together with related 
angular variables in the laboratory frame. While these variables are 
experimentally accessible and theoretically well motivated, 
their practical impact must be assessed in a realistic 
signal--background environment, where acceptance cuts, 
missing-energy reconstruction, and correlations among observables 
may dilute the true parton-level picture.
Moreover, these two polarization-sensitive observables provide powerful tools for discriminating between models that yield the same final state but arise from different underlying dynamics.
We present this study in three steps systematically. 
First, we discuss the underlying dynamics of ALP-like (V-V-a) and Higgs-like (V-V-H) associated production, emphasizing how the different coupling structures dictate gauge-boson polarization and recoil. Second, focusing on the decay of ALP to a pair of photons leading to 
final states with $\ell \gamma\gamma$ or $\ell\ell \gamma\gamma$, 
we perform a realistic signal--background simulation to demonstrate the characteristic features of gauge bosons produced in 
association with ALP and Higgs. Subsequently, using both cut-based selections and multivariate methods, including XGBoost, we predict the signal sensitivity for a given integrated luminosity option, 
say ${\cal L}=3~{\rm ab}^{-1}$, and then quantify the sensitivity to the ALP coupling scale with gauge bosons. 
This allows us to revisit the ALP analysis in the 
literature \cite{Cheung:2024qge} with an improved set of variables, while including a broader set of SM backgrounds.  
Third, we propose a strategy using polarization-sensitive variables to discriminate ALP signals from Higgs-like scenarios, as well as from simple BSM scalar and pseudoscalar toy models. This comparison isolates features specific to the ALP field-strength interaction which are more general consequences of spin, CP, and momentum-dependent electroweak couplings. 
Thus, the associated gauge boson is used not only as an event tag, but also as a diagnostic handle for interpreting a possible spin--0 signal in $V+X$ final states.

This paper is organized as follows. The required theoretical setup is presented in Sec.~\ref{sec:theory_setup}. 
Sec.~\ref{sec:simulation_observables} provides a description of Monte Carlo simulation of signal and background and construction of observables.
In Sec.~\ref{sec:analysis}, the estimation of signal and backgrounds is presented, including the signal and ALP-gauge boson coupling sensitivities for the HL-LHC option. The methodology of model discrimination is presented in Sec.~\ref{sec:model_disc}. Finally, we summarize and conclude in Sec.~\ref{sec:conclusions}. 
\section{Theoretical Framework}
\label{sec:theory_setup}

Axion-like particles (ALPs) are pseudo-Nambu--Goldstone bosons (pNGB) associated with an approximate softly broken shift symmetry. The general interactions of ALPs with SM fields can be described within the framework of effective field theory (EFT).  At leading order, dimension-five operators describe the relevant interactions in terms of the ALP's coupling to gauge bosons, along with separate derivative operators for its coupling to fermions. The general interaction Lagrangian is given by
\cite{GEORGI198673},
\begin{align}
{\cal L} &=  \frac12(\partial_\mu a)(\partial^\mu a)-\frac12 m_a^2 a^2 \nonumber\\
&- \frac{C_{GG}}{f_a}a {G^a}_{\mu\nu}{\tilde G}^{a\mu\nu}
- \frac{C_{BB}}{f_a}a B_{\mu\nu}{\tilde B}^{\mu\nu}
- \frac{C_{WW}}{f_a}a {W^I}_{\mu\nu}{\tilde W}^{I\mu\nu}
- \frac{\partial^\mu a} {f_a} \sum_f \bar \psi_f C_f\gamma_\mu \psi_f
\label{eq:Lagalp}      
\end{align}
where $m_a$ and $f_a$ represent the mass and decay constant of the ALP, respectively. Here, the coefficients $C_i$ generically denote the Wilson coefficients of the corresponding operators, as indicated by their subscripts, with the interaction strengths suppressed by the ALP scale $f_a$. We assume that $f_a$ lies above the electroweak scale and that the kinematic regime probed in our analysis remains within the validity of the EFT description.
The gauge fields ${G^a}_{\mu\nu}, B_{\mu\nu}$ and ${W^I}_{\mu\nu}$ correspond to the gauge groups $\rm SU(3)_c,\, SU(2)_L$, and $\rm U(1)_Y$ respectively. 
The second term presents the mass term for the ALP.
The last term runs over all fermions in the SM.
In principle, $C_f$ is a $3\times3$ matrix in flavor space, incorporating mixing among the different flavors, which is diagonal in the absence of any mixing.  
The operators describing the couplings of the ALP to gauge fields are
invariant under a shift symmetry $a\to a+\alpha$, where $\alpha$ is a
constant, since their variation is proportional to
$V_{\mu\nu}\widetilde V^{\mu\nu}$, which is a total derivative.  In the case of operators describing the ALP couplings to fermions, the derivative coupling through $\partial_\mu a$ automatically preserves the shift symmetry. The ALP mass term, on the other hand, softly breaks
the continuous shift symmetry, allowing a generic ALP to acquire a mass which is not fixed by the usual QCD-axion mass--decay-constant relation
\cite{Holdom:1982ex,dimopoulos2016colliderobservableqcdaxion,Hook_2020}.
Note that in the dimension-five Lagrangian given in 
Eq.~\eqref{eq:Lagalp}, there is no coupling term involving the scalar field, such as
$\frac{\partial_\mu a}{f_a} \phi^\dagger 
i{\cal \overleftrightarrow D}_\mu \phi$.   
This operator can be absorbed into the corresponding fermionic operators by using the equations of motion for the Higgs and fermion fields~\cite{GEORGI198673}. We therefore work in a basis in which
this operator is absent. However, the interaction of 
ALP with Higgs can be written using dimension-six operators.

In this work, we concentrate on ALP production in association with electroweak
gauge bosons and therefore retain only those interactions in
Eq.~\eqref{eq:Lagalp}, setting $C_{GG}=C_f=0$.
Hence, we start with the leading ${\cal O}({1/f_a})$ Lagrangian,  
\begin{align}
\mathcal{L}_{aVV}
&=
-\frac{a}{f_a}
\Big[
C_{WW}\,W^i_{\mu\nu}\widetilde W^{i\mu\nu}
+C_{BB}\,B_{\mu\nu}\widetilde B^{\mu\nu}
\Big], \label{eq:LaVV}
\quad
{\rm with} \,\,\, \widetilde V^{\mu\nu}
\equiv
\frac{1}{2}\epsilon^{\mu\nu\rho\sigma}V_{\rho\sigma},\,\,\,
{\rm and } \,\,\, \epsilon^{0123}=+1,
\end{align}
which implies the characteristic three-point a-V-V interaction,
\begin{align}
i\mathcal{V}^{\alpha\beta}_{aVV}(q_1,q_2)
=
-\,i\,\frac{4C_{VV}}{f_a}\,
\epsilon^{\alpha\beta\rho\sigma}q_{1\rho}q_{2\sigma},
\label{eq:aVV_vertex}
\end{align}
where $q_1$ and $q_2$ represent the momenta of two gauge bosons involved in the vertex.
For comparison, we consider a Higgs-like CP-even scalar with an SM-like non-derivative coupling to the massive electroweak gauge bosons, which, at the level of the three-point vertex, can be written as,
\begin{align}
i\mathcal{V}^{\alpha\beta}_{HVV}
=
i\,g_{HVV}\,g^{\alpha\beta},
\qquad
g_{HVV}=\frac{2m_V^2}{v},
\label{eq:HVV_vertex}
\end{align}
where \(v=(\sqrt{2}G_F)^{-1/2}\simeq246~{\rm GeV}\) is the electroweak vacuum
expectation value. 

After electroweak symmetry breaking, Eq.~\eqref{eq:LaVV} can be rearranged in terms of the physical basis,
\begin{align}
\mathcal{L}_{aVV}
=
-\frac{a}{4}
\bigg[
&
g_{a\gamma\gamma}
F_{\mu\nu}\widetilde F^{\mu\nu}
+
g_{aZZ}
Z_{\mu\nu}\widetilde Z^{\mu\nu}
\nonumber \\
&+
g_{aZ\gamma}
F_{\mu\nu}\widetilde Z^{\mu\nu}
+
2\, g_{aWW}\,W^+_{\mu\nu}\widetilde W^{-\,\mu\nu}
\bigg].
\label{eq:alp_mass_basis}
\end{align}
Using $s_w=\sin\theta_W$,$c_w=\cos\theta_W$, the effective dimensionfull 
ALP-gauge boson
couplings can be redefined as,      
\begin{align}
g_{a\gamma\gamma}
&=
\frac{4}{f_a}
\left(C_{BB}c_w^2+C_{WW}s_w^2\right),
&
g_{aWW}
&=
\frac{4C_{WW}}{f_a},
\nonumber\\
g_{aZZ}
&=
\frac{4}{f_a}
\left(C_{BB}s_w^2+C_{WW}c_w^2\right),
&
g_{aZ\gamma}
&=
\frac{8s_wc_w}{f_a}
\left(C_{WW}-C_{BB}\right).
\label{eq:alp_couplings}
\end{align}
Let us now closely examine the $V$-$V$-$a$ vertex in Eq.~\eqref{eq:aVV_vertex}, which determines the polarization pattern of the final-state gauge boson.
This behavior follows directly from the antisymmetric Lorentz structure
of the vertex as explained below.

For definiteness, let us consider the
partonic process of ALP production via s-channel mediated by a gauge boson, 
\begin{align}
q\bar q^{(\prime)}
\to
V_i^\ast(q)
\to
V_f(p_V,\lambda)\,a(p_a),
\qquad
q^\mu=p_V^\mu+p_a^\mu,
\label{eq:partonic_ALP_polarization}
\end{align}
where $V_f=W^\pm,Z$.  For $W^\pm a$ production, $V_i=W^\pm$, while for
$Za$ production both $V_i=Z$ and $\gamma$ contribute.
The part of the amplitude relevant for the final-state gauge-boson helicity can be expressed as,
\begin{align}
\mathcal{M}^{(a)}_\lambda
\propto
J^\mu\,\Delta_{\mu\alpha}(q)\,
H^{\alpha\beta}(q,p_V)\,
\varepsilon_\beta^\ast(p_V,\lambda), \qquad \text{with } H^{\alpha\beta}(q,p_V)
\equiv
\epsilon^{\alpha\beta\rho\sigma}q_\rho p_{V\sigma}.
\label{eq:amp_ALP_polarization}
\end{align}
Here $J^\mu$ denotes the initial-state quark current,
$\Delta^{(i)}_{\mu\alpha}(q)$ is the propagator of the intermediate gauge
boson $V_i$, and $\varepsilon^\beta(p_V,\lambda)$ is the polarization vector
of the on-shell final-state gauge boson.

It is important to distinguish two different consequences of the ALP tensor
structure.  First, any propagator term proportional to the momentum $q_\alpha$
on the ALP-vertex index gives no contribution,i.e.
\begin{align}
q_\alpha H^{\alpha\beta}(q,p_V)=q_\alpha\epsilon^{\alpha\beta\rho\sigma}q_\rho p_{V\sigma}=0 .
\label{eq:propagator_longitudinal_vanishes}
\end{align}
This follows because $q_\alpha q_\rho$ is symmetric under
$\alpha\leftrightarrow\rho$, whereas the Levi-Civita tensor is
antisymmetric.  Naturally, the longitudinal momentum-dependent term in the intermediate gauge-boson propagator does not contribute to the amplitude,
including the gauge-dependent term that may appear in a general gauge.
Note that this result by itself does not determine the physical
polarization of the final-state $W$ or $Z$ boson.
 
The physical helicity selection follows from a separate contraction with the
on-shell polarization vector of the final gauge boson. In $2\to2$ kinematics, the two independent four-vectors $q$ and $p_V$ define the production plane. Owing to the antisymmetric
Levi-Civita tensor, $H^{\alpha\beta}$ is orthogonal to both $q$ and $p_V$.
This becomes transparent in the partonic centre-of-mass frame,
with the $z$-axis chosen along the direction of $\vec p_V$,
\begin{align}
q^\alpha
&=
(\sqrt{\hat s},0,0,0),
\qquad
p_V^\alpha
=
(E_V,0,0,p),
\\ \nonumber
&\implies
H^{\alpha\beta}
=
\sqrt{\hat s}\,p\,\epsilon^{\alpha\beta 0 3}.
\label{eq:H_CM}
\end{align}
This implies $H^{\alpha 0}=H^{\alpha 3}=0$ for all $\alpha$. Since the longitudinal polarization vector of the final-state gauge boson is
\begin{align}
\varepsilon^\alpha(p_V,0)
=
\frac{1}{m_V}(p,0,0,E_V),
\end{align}
and therefore has support only in the $(0,3)$ components, its contraction
with $H^{\alpha\beta}$ vanishes, i.e., $H^{\alpha 0}=H^{\alpha 3}=0$.
Therefore, the longitudinal helicity
amplitude vanishes identically at tree level,
\begin{align}
\mathcal{M}^{(a)}_0(q\bar q^{(\prime)}\to Va)=0,
\qquad
\mathcal{M}^{(a)}_{\pm}\neq 0.
\label{eq:ALP_transverse}
\end{align}
Thus, for on-shell $V$ production along with an ALP,  $q\bar q\to Va$ at LO, 
the recoiling gauge boson is produced purely transversely in the $Va$ partonic center-of-mass helicity basis~\cite{Stong:1993bq,Bedi:2025hbz}. This is a
characteristic consequence of the antisymmetric Lorentz structure of the
interaction in Eq.~\eqref{eq:aVV_vertex} and provides a useful handle for
distinguishing ALP production from other spin--0 scenarios, as discussed in
the following sections.

In contrast, for a Higgs-like scalar with the vertex in
Eq.~\eqref{eq:HVV_vertex}, the antisymmetric projection responsible for
Eq.~\eqref{eq:ALP_transverse} is absent, and hence the 
longitudinal component of the final-state gauge-boson amplitude does not vanish, allowing both longitudinal and transverse modes to contribute \cite{Nakamura:2017ihk, Rao:2018abz, Rao:2020hel, Colyer:2025ehv}. 
However, in the boosted regime, with large transverse momentum of the gauge boson, the longitudinal polarization can become particularly important in SM Higgs production. This behavior can be understood through the Goldstone Boson Equivalence Theorem, according to which, at energies much larger than the gauge-boson mass, the longitudinal polarization of a massive gauge boson is effectively described by the corresponding Goldstone boson \cite{Cornwall:1974km,Lee:1977eg,Chanowitz:1985hj}. In this limit, the
longitudinal polarization vector behaves as
\begin{align}
\varepsilon^\mu(p_V,0)
=
\frac{p_V^\mu}{m_V}
+
\mathcal{O}\left(\frac{m_V}{E_V}\right).
\end{align}
This contrast is one of the key observations for using polarization-sensitive observables in the analysis. 

The same $aVV$ tensor structure that determines the polarization of the
associated gauge boson also leads to characteristic recoil kinematics.
In the following, we elaborate on these complementary features and introduce
two classes of observables: first those probing the polarization pattern
discussed above, and subsequently those probing the recoil kinematics.

\paragraph{Polarimetry from leptonic decays:}
As discussed above, the gauge boson produced in association with an ALP is
predominantly transversely polarized as a consequence of the tensor structure
of the $aVV$ interaction.
Experimentally, the polarization of $W$ and $Z$ bosons has been measured in
several processes~\cite{CMS:2011kaj,ATLAS:2012au,CDF:2010cnn,
ATLAS:2019bsc,CMS:2021icx}. In practice, the polarization information can be extracted
from the angular distributions and correlations of the gauge-boson decay
products, in particular the charged leptons. In the helicity frame, the charged lepton is viewed in the rest frame of the
gauge boson, with the quantization axis chosen along the direction of motion
of the gauge boson in the laboratory frame. In this frame, an angle $\theta^\ast_{\ell V}$ is defined as the polar angle of the charged lepton (in this case $\ell^+$) with respect to the $W^+$ or $Z$ flight direction observed in the laboratory frame, as given by,
\begin{align}
	\cos\theta^\ast_{\ell V}= \hat{\mathbf p}^{\,*}_{\ell}\cdot \hat{\mathbf p}^{\,{\rm lab}}_{V}.
	\label{eq:costheta}
\end{align}
For the leptonic decay of a $W^+$ boson, after integrating over the azimuthal
angle, the normalized polar-angle distribution can be written as~\cite{Ellis:1996mzs,Bern:2011ie}, 

\begin{align}
\frac{1}{\sigma}
\frac{d\sigma}{d\cos\theta^\ast_{\ell V}}
=
\frac{3}{8}F_L(1 - \cos\theta^\ast_{\ell V})^2 
+
\frac{3}{8}F_R(1 + \cos\theta^\ast_{\ell V})^2 
+
\frac{3}{4}F_0 \sin^2\theta^\ast_{\ell V},
\label{eq:costheta_relations}
\end{align}
where $F_0$, $F_L$ and $F_R$ denote the longitudinal, left-handed and
right-handed polarization fractions, respectively, with
$F_0+F_L+F_R=1$. For $W^-$ decay, the roles of $F_L$ and $F_R$ are
interchanged for the same charged-lepton angular convention. Also note that Eq.~\eqref{eq:costheta_relations} applies to the ideal full phase space; fiducial
selection requirements can distort the angular distribution. For $W^+$ decay, using the normalized angular distribution together with
the normalization condition, the polarization fractions can be written as
\cite{Bern:2011ie},
\begin{equation}
\begin{aligned}
F_0 &= 2-5\left\langle\cos^2\theta^\ast_{\ell V}\right\rangle, \\
F_L &= -\frac{1}{2}-\left\langle\cos\theta^\ast_{\ell V}\right\rangle
      +\frac{5}{2}\left\langle\cos^2\theta^\ast_{\ell V}\right\rangle\\
F_R &= -\frac{1}{2}+\left\langle\cos\theta^\ast_{\ell V}\right\rangle
      +\frac{5}{2}\left\langle\cos^2\theta^\ast_{\ell V}\right\rangle .
\end{aligned}
\label{eq:polfrac}
\end{equation}
For the ALP production process considered here, as argued above in Eq.~\eqref{eq:ALP_transverse}, the longitudinal components of the gauge boson polarization are expected to be absent,
\begin{align}
F_0^{(a)}\simeq 0 ,
\end{align}
up to finite-width effects, QCD radiation, fiducial selections, and
reconstruction effects.  In contrast, a Higgs-like scalar can give a non-zero longitudinal fraction, $F_0^{H}\neq 0$.

\begin{figure}[t]
	\centering
		\centering
		\includegraphics[width=0.6\linewidth]{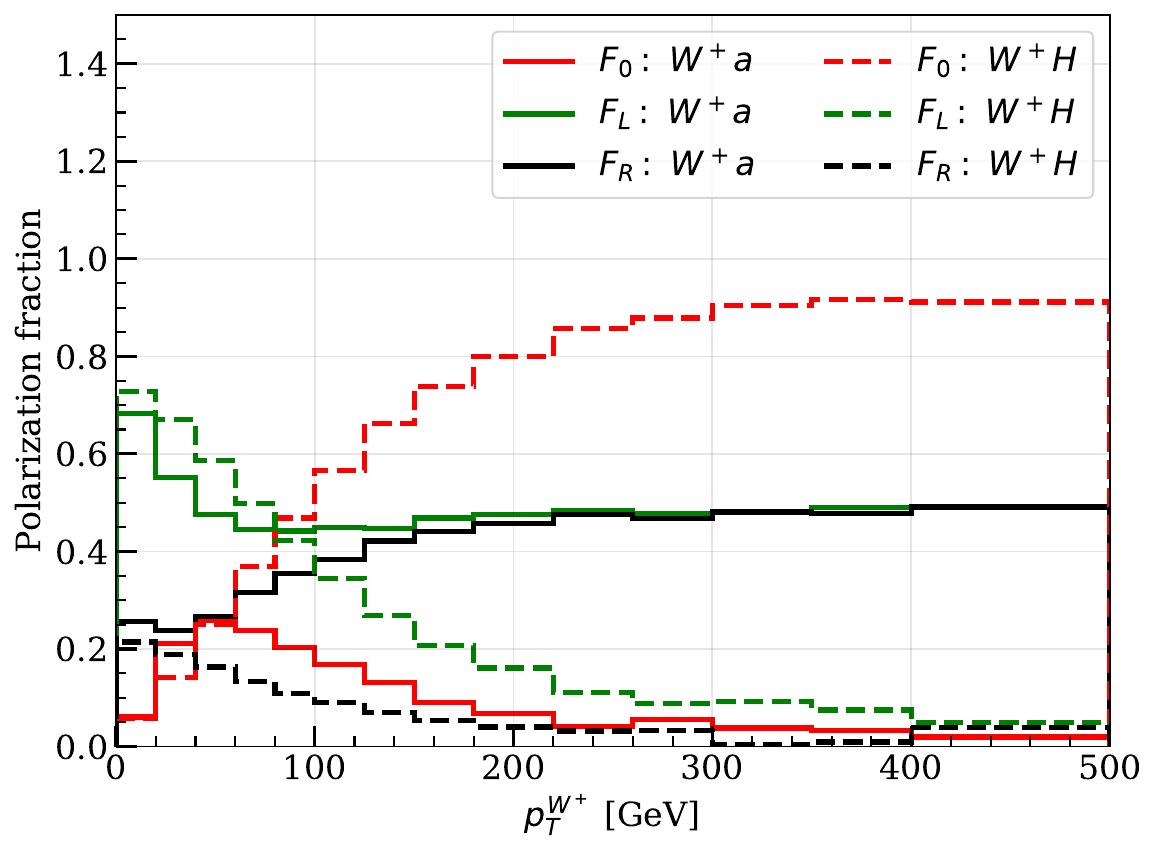}
	\caption{Generator-level $W$ polarization fractions $F_0, F_L$ and $F_R$ as a function of $p_T^W$ for $W^+a$ (solid) and $W^+ H$ (dashed) processes, setting $m_H = m_a$ = 125~GeV. No fiducial selections are applied.}
	\label{fig:polfrac_WH_ALP}
\end{figure}

This polarization pattern can be made explicit by studying the fractions
$F_0$, $F_L$ and $F_R$ as functions of the transverse momentum of the
associated gauge boson.  Fig.~\ref{fig:polfrac_WH_ALP} shows these polarization fractions for $W^+a$ and $W^+H$ production using truth-level information of final state particle momentum, without any fiducial selections. For the ALP case, the longitudinal fraction remains
strongly suppressed over the presented $p_T^W$ range, while the transverse
fractions dominate as discussed above. Whereas, in the Higgs-like case, the longitudinal component becomes increasingly important in the boosted regime, in accordance with the Goldstone-boson equivalence theorem discussed above. The difference between the ALP and Higgs-like cases is therefore directly reflected in the helicity-angle distributions.

This exercise demonstrates how the polarization information can be accessed through the $\cos\theta^\ast_{\ell V}$ distribution.
These parametric differences motivate the construction of polarization-sensitive observables to discriminate between the models and to suppress the SM backgrounds. The same $aVV$ tensor structure also leaves a characteristic imprint on the recoil kinematics of the associated gauge boson, which we discuss next.

\paragraph{Gauge-boson recoil momentum:}
The recoil momentum of the associated gauge boson in the 
ALP production, as shown in Eq.~\eqref{eq:partonic_ALP_polarization}, is found to be exceptionally hard. It can be attributed to the 
tensor structure of the $aVV$ vertex in Eq.~\eqref{eq:aVV_vertex}, which
introduces additional power of the hard momenta into the matrix element.
Consequently, high-recoil configurations are weighted differently from those
in a non-derivative Higgs-like interaction. 
Consider the generic process in Eq.~\eqref{eq:partonic_ALP_polarization}
in the partonic CM frame defined as, 
\begin{align}
q^\mu
&\equiv
(\sqrt{\hat s},0,0,0),
\\
p_V^\mu
&\equiv
(E_V,p\sin\theta,0,p\cos\theta),
\end{align}
where $p=|\vec p_V|$ and $\theta$ is the scattering angle of the final-state
gauge boson with respect to the beam direction.
Its transverse momentum is
therefore $p_T(V)=p\sin\theta$.
For the ALP interaction, only the transverse helicities of the final-state gauge boson contribute at tree level, as discussed above. For a single $s$-channel mediator $V_i$, the corresponding differential cross section scales as
\begin{align}
\frac{d\hat\sigma_a}{d\cos\theta}
\propto
\frac{g_{aVV}^{\,2}}{\sqrt{\hat s}\,D_i(\hat s)}
\,\hat s\,p^3\left(1+\cos^2\theta\right),
\end{align}
where $D_i(\hat s)$ is the squared denominator of the intermediate gauge-boson propagator.  For a Higgs-like scalar, both transverse and longitudinal gauge-boson polarizations contribute. In the boosted regime, the longitudinal contribution is enhanced and scales as
\begin{align}
\left.
\frac{d\hat\sigma_H}{d\cos\theta}
\right|_{L}
\propto
\frac{g_{HVV}^{\,2}}{\sqrt{\hat s}\,D_i(\hat s)}
\,p\,\frac{E_V^2}{m_V^2}\sin^2\theta
\;\sim\;
\frac{g_{HVV}^{\,2}}{\sqrt{\hat s}\,D_i(\hat s)}
\,\frac{p^3}{m_V^2}\sin^2\theta ,
\end{align}
where the last relation applies for $E_V\simeq p$ at high energy. 
In both cases, one power of $p$ originates from the two-body phase-space
factor.  The derivative $aV_{\mu\nu}\widetilde V^{\mu\nu}$ interaction
introduces an additional hard-momentum dependence, reflected by the factor
of $\hat s$ in the ALP cross section.  In the high-energy limit,
$p\simeq\sqrt{\hat s}/2$, so the ALP interaction preferentially weights
high-recoil configurations, leading to a harder normalized $p_T^V$ spectrum
than in the Higgs-like case.
This feature provides a useful handle for suppressing the SM backgrounds, including the associated SM Higgs production, 
enhancing the signal sensitivity.
It also plays an important role in the process of discriminating different
spin--0 production mechanisms showing identical signatures. 
In the subsequent sections, the recoil momentum distributions of the gauge boson for signal and background processes are presented, thereby validating this key observation. The harder recoil spectrum induced by the derivative ALP interaction indicates the expected energy growth of a dimension-five EFT amplitude, which does not by itself imply a breakdown of perturbative unitarity in the kinematic region
considered here. A recent complete partial-wave analysis finds that the
strongest constraints on the electroweak dimension-five gauge operators arise
from coupled bosonic channels with two operator insertions
\cite{Bresciani:2025ojh}. The individual bounds are
$\sqrt{\hat s}|C_{WW}/f_a|\lesssim 1.5$ and
$\sqrt{\hat s}|C_{BB}/f_a|\lesssim 1.9$.
For our nominal benchmarks, the combined $B$--$W$ scattering system reaches the perturbative-unitarity limit well above the partonic energies relevant for the LHC analysis.

\section{Monte Carlo Simulation and Observables}
\label{sec:simulation_observables}
\begin{figure}[t]
	\centering
	\begin{tikzpicture}
	\begin{feynman}
        \vertex (a) at (-1.5, 1.0) {$q$};
		\vertex (b) at (-1.5, -1.0) {$\bar q^{(\prime)}$};
		\vertex[dot] (e) at (-0.75, 0){};
		\vertex[blob, scale=0.5] (f) at (0.75, 0) {};
        \vertex[dot] (c) at (1.7, 0.6) {};
        \vertex (c1) at (2.8, 1.2) {$f_1$};
        \vertex (c2) at (2.8, 0.3) {$f_2$};
		\vertex[blob, scale=0.5] (d) at (1.7, -0.6) {};
        \vertex (d1) at (2.8, -1.2) {$\gamma$};
        \vertex (d2) at (2.8, -0.3) {$\gamma$};
		\diagram* {
			(a) -- [fermion] (e) -- [fermion] (b),
			(e) -- [photon, edge label=$V^\ast$] (f),
            (f) -- [photon, edge label=$V$] (c),
			(f) -- [scalar, edge label=$\phi$] (d),
            (c) -- [fermion] (c1),
            (c2) -- [fermion] (c),
            (d) -- [photon] (d1),
            (d2) -- [photon] (d),
		};
	\end{feynman}
	\end{tikzpicture}
	\caption{Representative Feynman diagram for associated production
	$q\bar q^{(\prime)}\to V^\ast\to V\phi$, followed by
	$\phi\to\gamma\gamma$ and the leptonic decay of the associated gauge boson. Here $\phi=a,S$, with $a$ denoting the ALP and $S$ denoting a spin--0 look-alike state.  The final-state fermions $f_1f_2$ denote
	$\ell^\pm\nu$ for $V=W^\pm$ and $\ell^+\ell^-$ for $V=Z$. The filled dots denote SM vertices, while the blobs denote
BSM vertices. }
	\label{fig:feynman_diag}
\end{figure}
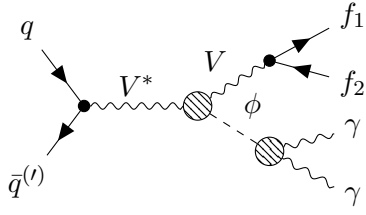

As mentioned above, we focus on the associated-production channels,
\begin{align}
pp&\to W^\pm(\to \ell^+\nu) \phi \\  
&\to Z(\to \ell^+\ell^-)\phi 
\label{eq:sim_processes}
\end{align}
with $\phi=a,S $,  
at $\sqrt{s}=14~{\rm TeV}$, followed by $\phi\to\gamma\gamma$, 
where S can be any scalar other than the ALP. Such light
scalar states can arise, for example, in singlet-extended Higgs sectors,
including the Next-to-Minimal Supersymmetric Standard Model (NMSSM)
\cite{FAYET1975104,Drees:1988fc,PhysRevD.39.844,Ellwanger:2009dp}
and 2HDM+singlet models
\cite{Branco:2011iw,Baum:2018zhf,Heinemeyer:2021msz}. These scenarios can give rise to the identical final state over the ALP mass range
considered here.  We discuss them in more detail in a later section in the
context of discriminating between different signal hypotheses.  A
representative topology for the associated-production process is shown in
Fig.~\ref{fig:feynman_diag}.

Events are generated with \texttt{MadGraph5\_aMC@NLO}\cite{Alwall:2014hca}, showered and hadronized with \texttt{PYTHIA8}\cite{Sjostrand:2007gs} and subsequently passed through \texttt{Delphes}\cite{deFavereau:2013fsa} using the CMS detector card to model detector effects. The \texttt{Monash 2013} tune \cite{Skands:2014pea} is used for ISR, FSR, and MPI effects. The renormalization scale ($\mu_R$) and factorization scale ($\mu_F$)
are set on an event-by-event basis to the partonic centre-of-mass energy ($\sqrt{\hat{s}}$).
The UFO\footnote{The ALP model is available at
\url{https://feynrules.irmp.ucl.ac.be/wiki/ALPsEFT}.}
implementation of the ALP effective theory~\cite{Brivio:2017ije}
is used to generate the signal process in Eq.~\eqref{eq:sim_processes}. 

We choose an illustrative electroweak benchmark setting gluon and
fermion couplings to zero, such that the ALP production and decay are controlled
primarily by the $aVV$ interactions.  This benchmark serves
as a reference point for generating the signal samples and for subsequently
rescaling the projected sensitivity. Existing bounds do not uniquely
determine such a benchmark, since the four mass-basis couplings are correlated
through the two independent combinations $C_{WW}/f_a$ and $C_{BB}/f_a$,
while experimental constraints are generally derived under specific
assumptions about the other couplings.  We therefore choose a representative
point consistent with these constraints under the coupling assumptions adopted
in Eq.~\eqref{eq:alp_couplings}.
Unless otherwise stated, we set~\cite{Carra:2021ycg, Gavela:2019cmq}
\begin{align}
f_a=1~{\rm TeV},\qquad C_{WW}=0.02,
\qquad C_{BB}=0.05,\qquad
C_{GG}=C_{a\Phi}=0.
\label{eq:alp_benchmark}
\end{align}
For convenience, from now on we absorb $1/f_a$ into the effective coupling and denote $C_{VV}/f_a$ by $C_{VV}$, with $C_{VV}$ expressed in units of $\mathrm{TeV}^{-1}$.  For reference, this corresponds to mass-basis couplings
\begin{align}
g_{a\gamma\gamma}
&\simeq
{\rm 1.71 \times 10^{-1} ~TeV}^{-1},
&
g_{aZ\gamma}
&\simeq
{\rm -1.01 \times 10^{-1} ~TeV}^{-1},
\nonumber\\
g_{aZZ}
&\simeq
{\rm 1.08 \times 10^{-1} ~TeV}^{-1},
&
g_{aWW}
&\simeq
{\rm 8 \times 10^{-2} \;TeV}^{-1}.
\end{align}
This represents a conservative benchmark choice consistent with existing
constraints~\cite{Carra:2021ycg, Gavela:2019cmq}. For comparison, we also present results for a more optimistic
benchmark, which remains allowed by the constraints reported in
Ref.~\cite{Carra:2021ycg, Gavela:2019cmq} with $(C_{WW}, C_{BB})$ parameter values $(0.09, 0.3)~\mathrm{TeV}^{-1}$. This benchmark point should not be interpreted as the result of a global fit
to existing constraints; rather, it provides a reference signal sample whose
normalization can subsequently be rescaled to determine the projected
sensitivity to the electroweak ALP couplings.

The ALP mass range considered in this study is
\begin{align}
20~{\rm GeV}\leq m_a\leq 200~{\rm GeV}.
\end{align}
The lower end of the mass range is chosen so that the two photons from
$a\to\gamma\gamma$ are typically reconstructed as resolved photons. 
Otherwise, for lighter ALPs, since the
opening angle scales approximately as
$\Delta R_{\gamma\gamma}\sim 2m_a/p_T(a)$, the two photons can become
collimated, requiring boosted-diphoton or photon-jet strategies
\cite{Wang:2021uyb,ATLAS:2022abz}. 
The total width and branching ratios of ALP are computed 
internally with
\texttt{MadGraph5\_aMC@NLO} for the benchmark in Eq.~\eqref{eq:alp_benchmark}.
In the absence of fermion and gluon couplings, the diphoton branching ratio of ALP decay is
close to unity. 
Once the $Z\gamma$, $W^+W^-$, and $ZZ$ decay channels become kinematically accessible, ${\rm BR}(a\to\gamma\gamma)$ decreases in a benchmark-dependent manner, since the corresponding partial widths are controlled by the electroweak coupling combinations given in Eq.~\eqref{eq:alp_couplings}.
For instance, for the considered benchmark values of couplings 
Eq.~\eqref{eq:alp_benchmark}, the branching fraction to a pair of gauge bosons($WW$,$ZZ$, $Z\gamma$) are found to be less than 1\% for 
$m_a < 100$~GeV.  For a higher mass range, e.g 
at $m_a$=200~{GeV} the relevant branching fractions are, 
\begin{align}
{\rm BR}&(a\to\gamma\gamma)\simeq0.83,~~\quad
{\rm BR}(a\to W^+W^-)\simeq0.078,\nonumber\\
{\rm BR}&(a\to Z\gamma)\simeq0.072,\quad
{\rm BR}(a\to ZZ)\simeq0.023.
\end{align}
For the $\ell\gamma\gamma$ and $\ell\ell\gamma\gamma$ signal channels, the
dominant SM backgrounds are due to the 
\begin{align}
pp &\to W^\pm H,\;\;W^\pm\gamma\gamma,\;\; W^\pm\gamma j, \;\; Z\gamma \gamma 
\quad {\rm for}\quad Wa
\nonumber\\
&\to ZH,\;\;Z\gamma\gamma,\;\; Z\gamma j \quad {\rm for}\quad  Za
\end{align}
While $V\gamma\gamma$ processes constitute irreducible backgrounds, in the case of $V\gamma j$, one of the reconstructed photons originates from a jet misidentified as a photon.  We include this contribution using a constant jet-to-photon fake rate ${\rm FR}(j\to\gamma)=5\times10^{-4}$ \cite{ATL-PHYS-PUB-2009-005}.
The $Z\gamma\gamma$ process also contributes to the single-lepton channel when
one of the leptons from the $Z$ decay is not reconstructed, either because it
fails the lepton-selection requirements or lies outside the detector
acceptance.  This lost-lepton contribution is included in the background to ensure a realistic estimate of the signal sensitivity. 
In addition, associated SM Higgs production,
$pp\to VH$ followed by $H\to\gamma\gamma$, constitutes a resonant background,
particularly it becomes important for ALP masses close to the SM Higgs mass, and is therefore
included. 
The continuum backgrounds are normalized to the leading-order \texttt{MadGraph5\_aMC@NLO} cross sections.

While generating the events for analysis, we impose the following basic parton-level
requirements:
\begin{align}
p_T^\gamma>10~{\rm GeV},\quad
p_T^{\ell_{1(2)}}>20(10)~{\rm GeV},\quad
|\eta_{\gamma,\ell}|<2.5,\quad
\Delta R(i,j)>0.3; ~~i,j=\gamma,\ell,
\label{eq:gencuts}
\end{align}
where $\ell_1$ and $\ell_2$ correspond to leading and sub-leading leptons.
We now construct the polarization sensitive kinematics observables, following the motivation developed in the previous section.

\paragraph{$\cos\theta^\ast_{\ell V}$:} This variable is already defined in the previous section in Eq.~\eqref{eq:costheta}.
This angular variable probes the polarization structure of
the associated gauge boson through the angular distribution of its leptonic
decay products. Similar observables have been widely used in studies of
gauge-boson polarization in the SM~\cite{Bern:2011ie,Stirling:2012zt}, and we
adopt the same general strategy here.
The reconstructed momentum of the gauge boson is used to boost the charged
lepton into the gauge-boson rest frame.
For the $W$ boson, however, the reconstruction of its momentum in the leptonic decay channel is non-trivial because of the unobserved neutrino. 
By imposing the $W$-mass constraint on the lepton--neutrino system, the longitudinal component of the neutrino momentum can, in principle, be reconstructed, although the solution is not always unique and may require an additional prescription.  Such criteria may, however, reduce the signal acceptance and hence the overall efficiency of the analysis.
The strategy
adopted in our simulation for reconstructing the $W$-boson momentum is
described in Appendix~\ref{sec:App_W_momenta}.  In contrast, for
$Z\to\ell^+\ell^-$, the $Z$-boson momentum can be reconstructed directly from
the two charged leptons without such an ambiguity.

Alternatively, the CMS Collaboration employed the lepton-projection variable
$L_p$ in a study of boosted $W$-boson polarization~\cite{CMS:2011kaj},
defined as
\begin{align}
L_p
=
\frac{\vec p_T^\ell\cdot \vec p_T^W}{|\vec p_T^W|^2}.
\label{eq:Lp}
\end{align}
Since this variable depends only on transverse momenta, it avoids the
ambiguity associated with reconstructing the longitudinal momentum of the
neutrino. 
At large $p_T(W)$, it can be shown that $\cos\theta_{\ell V}^\ast = 2( L_p-1/2)$, implying a strong correlation between $\cos\theta_{\ell V}^\ast$ and $L_p$. 

As an alternative for reconstructing the full $W$ momentum, required for Eq.~\eqref{eq:costheta}, we use the di-photon system as a proxy for the recoil direction of the $W$ boson.
It is experimentally advantageous, since photons are cleanly measured objects and their momenta can be reconstructed with high precision \cite{CMS:2015myp, CMS:2020uim}
The charged lepton is boosted to the $W$ rest frame using the reconstructed W boson momentum, while the reference axis is taken to be the laboratory-frame direction of the di-photon system. We therefore introduce
the modified angular variable
\begin{align}
\cos\theta^\ast_{\ell\gamma\gamma}
=
\hat{\mathbf p}^{\,*}_{\ell}\cdot
\hat{\mathbf p}^{\,{\rm lab}}_{\gamma\gamma}.
\label{eq:costheta_gammagamma_def}
\end{align}
Although $\cos\theta^\ast_{\ell\gamma\gamma}$ is not identical to the true
helicity angle $\cos\theta^\ast_{\ell V}$, it also plays a role in capturing the similar gauge boson polarization effect.

\begin{figure*}[t]
	\centering
	\begin{minipage}{0.49\textwidth}
		\centering
		\includegraphics[width=\linewidth]{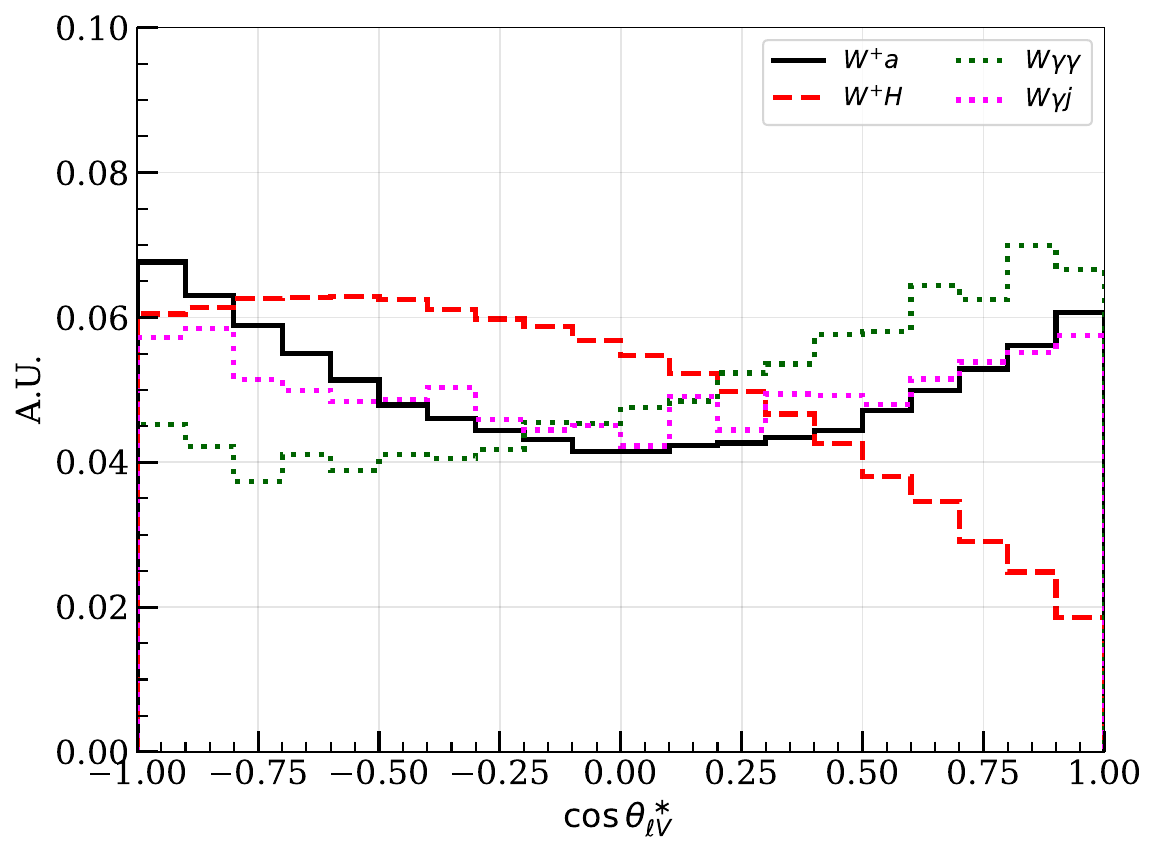}
		\vspace{1mm}
	\end{minipage}
	\hfill
	\begin{minipage}{0.49\textwidth}
		\centering
		\includegraphics[width=\linewidth]{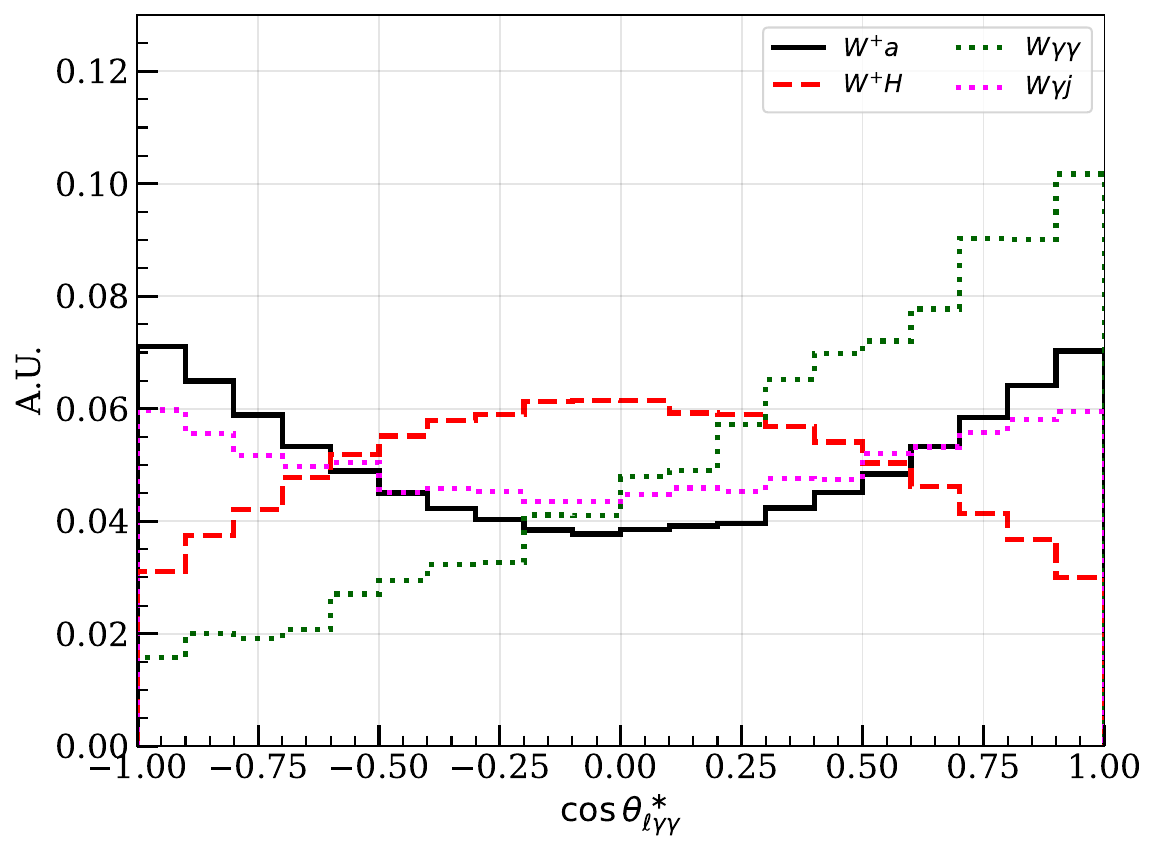}
		\vspace{1mm}
	\end{minipage}
	\caption{Normalized generator-level distributions of
$\cos\theta^\ast_{\ell V}$ (left) and
$\cos\theta^\ast_{\ell\gamma\gamma}$ (right) for
$W^+a$, $W^+H$, $W\gamma\gamma$, and $W\gamma j$ production, with
$m_a=m_H=125~{\rm GeV}$. No selection is applied other than
$p_T^W>50~{\rm GeV}$. The distributions illustrate the different
polarization patterns of the ALP and Higgs-like signals and the corresponding
continuum backgrounds.}
	\label{fig:costheta_gen_reco}
\end{figure*}

In Fig.~\ref{fig:costheta_gen_reco}, the distributions of $\cos\theta_{\ell V}^\ast$ (left) and $\cos\theta_{\ell \gamma \gamma}^\ast$ (right), are presented at the particle(truth)level   
without imposing any selection requirements other than 
$p_T(W)> 50$ GeV. The corresponding distributions for the SM backgrounds, including $WH$ processes, are also presented for the sake of comparison. For the ALP production case, the $W$ boson is predominantly transversely polarized, with almost equal contributions of the left- and right-handed polarization states. This is reflected in the characteristic enhancement of the angular
distribution near its endpoints ($\sim$ -1 or 1). 
For $W^+$ decay, the right-handed polarization component peaks near
$\cos\theta_{\ell V}^\ast\simeq+1$ 
($\theta_{\ell V}^\ast\simeq0$), whereas the left-handed component peaks
near $\cos\theta_{\ell V}^\ast\simeq-1$
($\theta_{\ell V}^\ast\simeq\pi$).  This behavior follows from angular-momentum
conservation together with the chiral structure of the $W\ell\nu$ interaction.
The corresponding pattern for $\cos\theta_{\ell\gamma\gamma}^\ast$ are similar. For the Higgs-like process, the polarization composition of the associated
$W^+$ differs significantly from that in ALP production.  In the
$\cos\theta^\ast_{\ell V}$ distribution, the sizable left-handed component
enhances the region at negative $\cos\theta^\ast_{\ell V}$, while the
longitudinal contribution populates the central region around
$\cos\theta^\ast_{\ell V}\simeq 0$.  Consequently, the distribution is more
asymmetric than in the ALP case, where the transverse left- and right-handed
components are more comparable. A similar distinction is visible in the recoil-axis variable
$\cos\theta^\ast_{\ell\gamma\gamma}$.  In this case, the Higgs sample
shows a more pronounced enhancement in the central region, reflecting its
larger longitudinal-polarization component, whereas the ALP distribution
retains a stronger endpoint structure associated with the dominance of
transverse polarizations.  The different orientation of the diphoton recoil
axis relative to the $W$ flight direction also modifies the shape
with respect to the conventional helicity-angle distribution.

The continuum backgrounds exhibit different angular patterns.  In
particular, the $W\gamma\gamma$ contribution is enhanced toward one endpoint
of the distribution, indicating a different helicity composition from both
the ALP and Higgs-like signals, while the $W\gamma j$ background is
comparatively less structured.  These differences further motivate the use
of the angular observables in the subsequent signal--background analysis.

\begin{figure*}[t]
	\centering
	\begin{minipage}{0.49\textwidth}
		\centering
	\includegraphics[width=\linewidth]{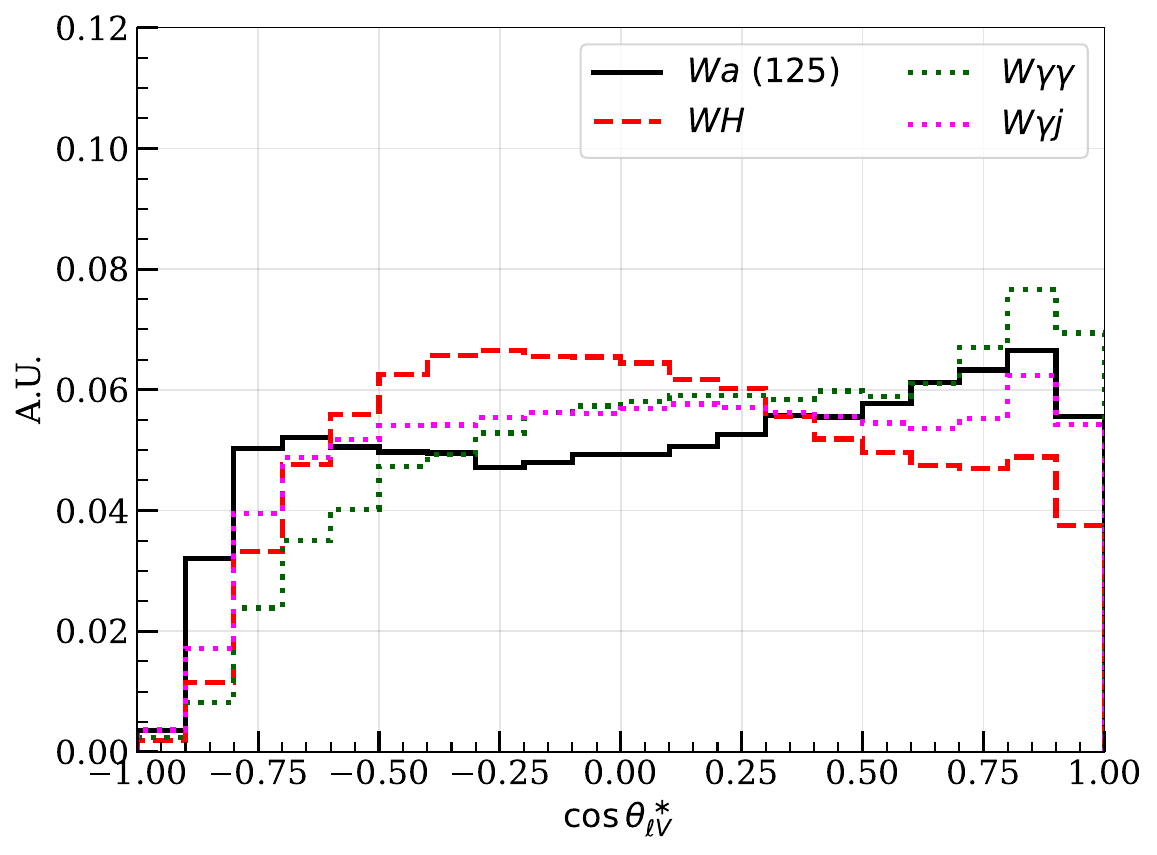}
		\vspace{1mm}
	\end{minipage}
	\hfill
	\begin{minipage}{0.49\textwidth}
		\centering
		\includegraphics[width=\linewidth]
{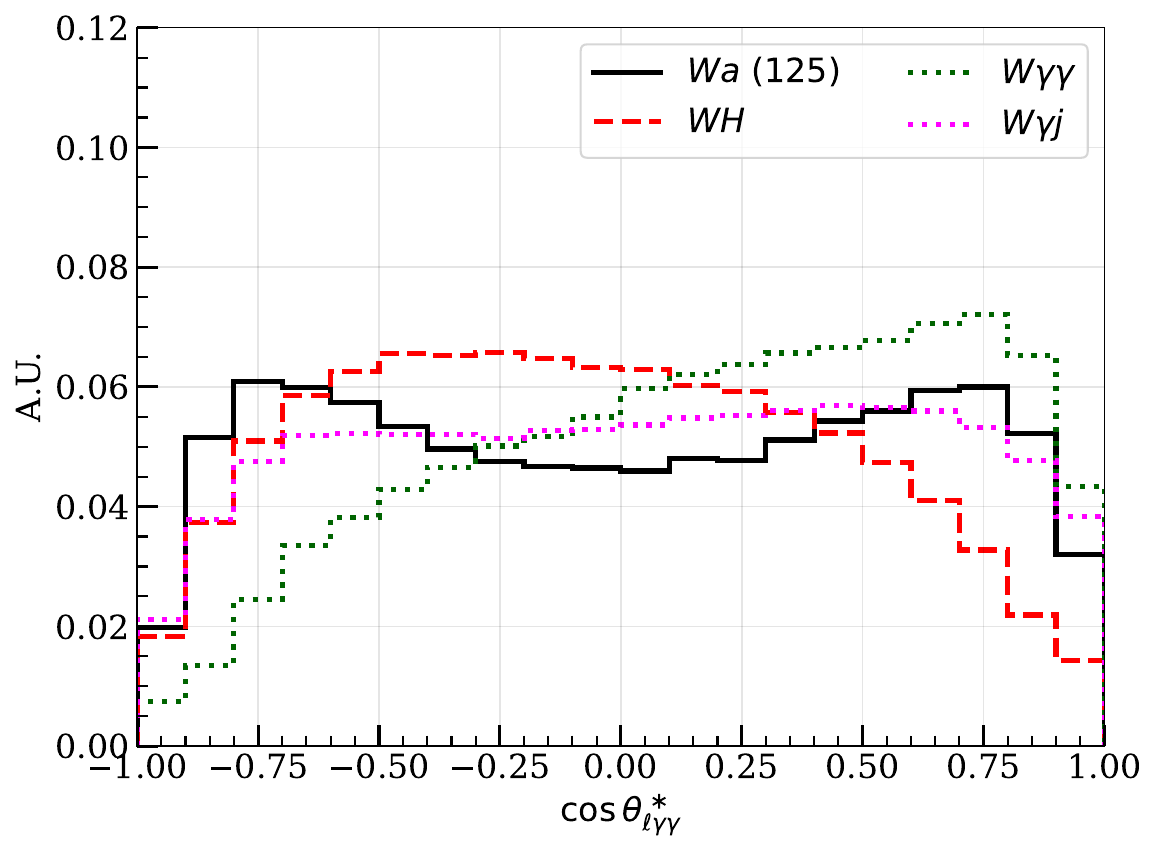}
		\vspace{1mm}
	\end{minipage}
	\caption{Same as Fig.~\ref{fig:costheta_gen_reco}, but at reconstruction level after
applying the baseline selections in Eq.~\eqref{eq:gencuts}, together with
$p_T^W>50~{\rm GeV}$ and $\met>15~{\rm GeV}$. The lepton-$p_T$ and $\met$ requirement distorts the distributions near the endpoints, while the
central region is comparatively less affected.}
	\label{fig:costheta_reco}
\end{figure*}

These distributions clearly demonstrate distinctive features of the $W$ boson polarization in processes involving ALP and Higgs. Notably,
the characteristic features become even more pronounced in 
the boosted regime, where the $W$ boson carries higher transverse momentum. However, this clear distinction is significantly affected once selection requirements are imposed on the final state leptons at the reconstruction level. Consequently, the shapes of the angular distributions become distorted, particularly in the regions around the endpoints, as shown in
Fig.~\ref{fig:costheta_reco}, which are subjected to selections in Eq.~\eqref{eq:gencuts} along with $\met>$15~GeV. The origin of this effect can be realized from the following arguments. In the $W$ rest frame,
$E_\ell^\ast=E_\nu^\ast=m_W/2$, and a boost($\gamma \sim \frac{E}{m_W}$) gives,
\begin{align}
E_\ell
&=
\gamma\,\frac{m_W}{2}\,(1+\beta\cos\theta^\ast),
&
E_\nu
&=
\gamma\,\frac{m_W}{2}\,(1-\beta\cos\theta^\ast).
\label{eq:lepton_neutrino_boost}
\end{align}
Thus, for the chosen convention, the charged lepton becomes soft near
$\cos\theta^\ast_{\ell\ell V/\ell \gamma\gamma} \simeq -1$, while the neutrino becomes soft near
$\cos\theta^\ast_{\ell\ell V/\ell \gamma\gamma}=+1$.
Lepton-$p_T$ and $\met$ selection cuts therefore
affect the endpoints of the reconstructed angular distribution.  
The middle of the distribution is less affected, because both factors $(1\pm\beta\cos\theta^\ast)$ 
remain of order unity for $\cos\theta^\ast\simeq 0$. This endpoint-clipping explains the main difference between the generator-level and reconstructed-level $\cos\theta^\ast$ shapes. 

In this context, it is worth emphasizing the importance of these angular distributions.  In experimental analyses, selection criteria must be imposed to reconstruct and identify physical objects using the information provided by different detector components. These criteria typically include kinematic requirements, such as minimum transverse-momentum thresholds on the reconstructed objects,
which can significantly reduce the sensitivity to the characteristic polarization features discussed above. However, if such thresholds could be lowered without compromising the reconstruction
performance, a larger fraction of the underlying polarization information
could potentially be retained. For instance, if low-level experimental inputs, such as calorimeter energy deposits and track momenta, etc. \cite{CMS:2017yfk} are used to reconstruct objects with minimal selection, then the underlying polarization information can be better recovered. In this context, the Machine-Learning-based approaches \cite{Searcy:2015apa, Grossi:2020orx, Bhattacharya:2020aid, Barman:2021yfh} may provide a promising framework for extracting polarization-sensitive information, particularly when combined with low-level detector inputs
\cite{Bols:2020bkb,Bhimji:2017qvb,Andrews:2018nwy}. Exploring such machine-learning-based reconstruction strategies is, however, beyond the scope of the present study and is left for future work.

Another production-angle observable, useful for probing the tensor structure of
the spin--0 interaction, is
\begin{align}
\cos\beta^\ast
=
\hat{\mathbf p}^{V\phi}_{V}\cdot \hat{\mathbf z},
\label{eq:betarstar}
\end{align}
where $\hat{\mathbf p}^{\,V\phi}_{V}$ denotes the direction of the vector boson in the $V\phi$ centre-of-mass frame and $\hat{\mathbf z}$ is the beam direction. 
Thus angle $\beta^\ast$ characterizes the production geometry of the $V\phi$ system,
rather than the decay of the gauge boson.
Its distribution is therefore sensitive to the different Lorentz
structures of the $HVV$ and $aVV$ interactions.

\begin{figure}[t]
	\centering
	\begin{minipage}{0.48\textwidth}
		\centering
		\includegraphics[width=\linewidth]{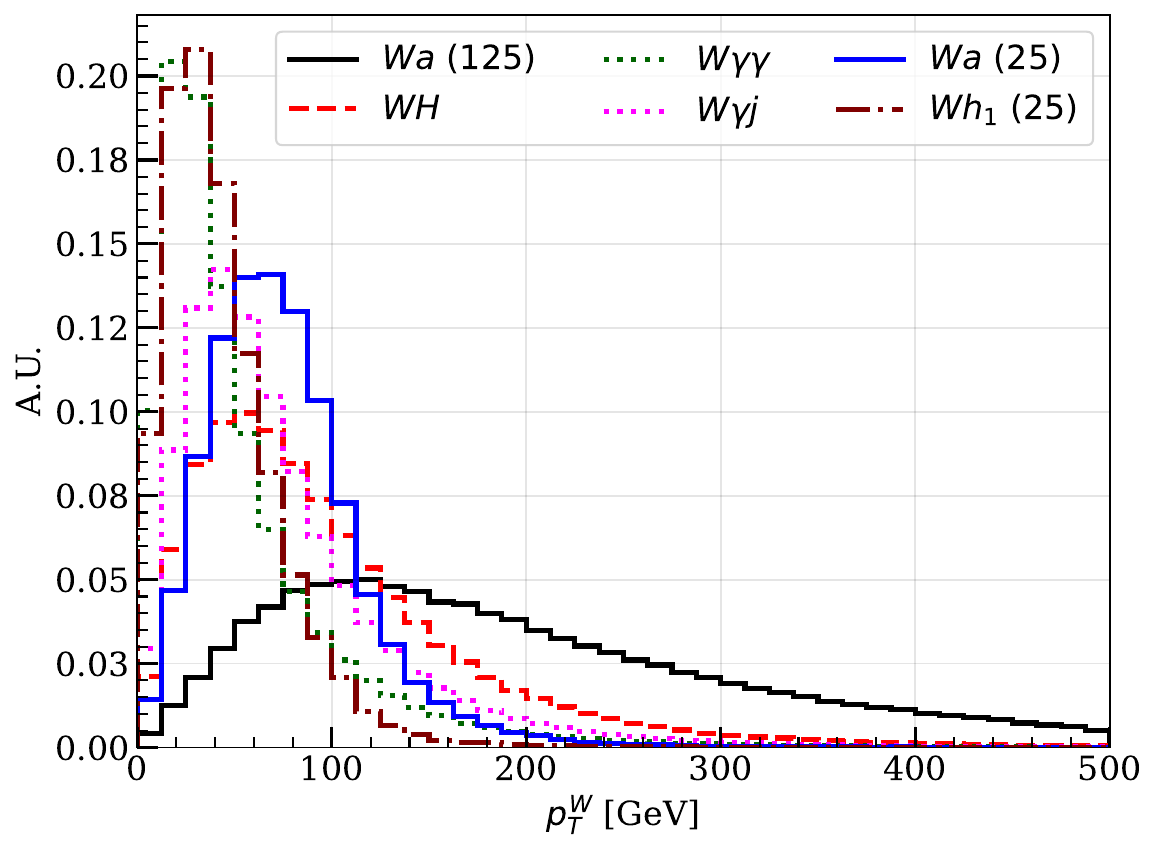}
		\vspace{1mm}
	\end{minipage}
	\hfill
	\begin{minipage}{0.48\textwidth}
		\centering
		\includegraphics[width=\linewidth]{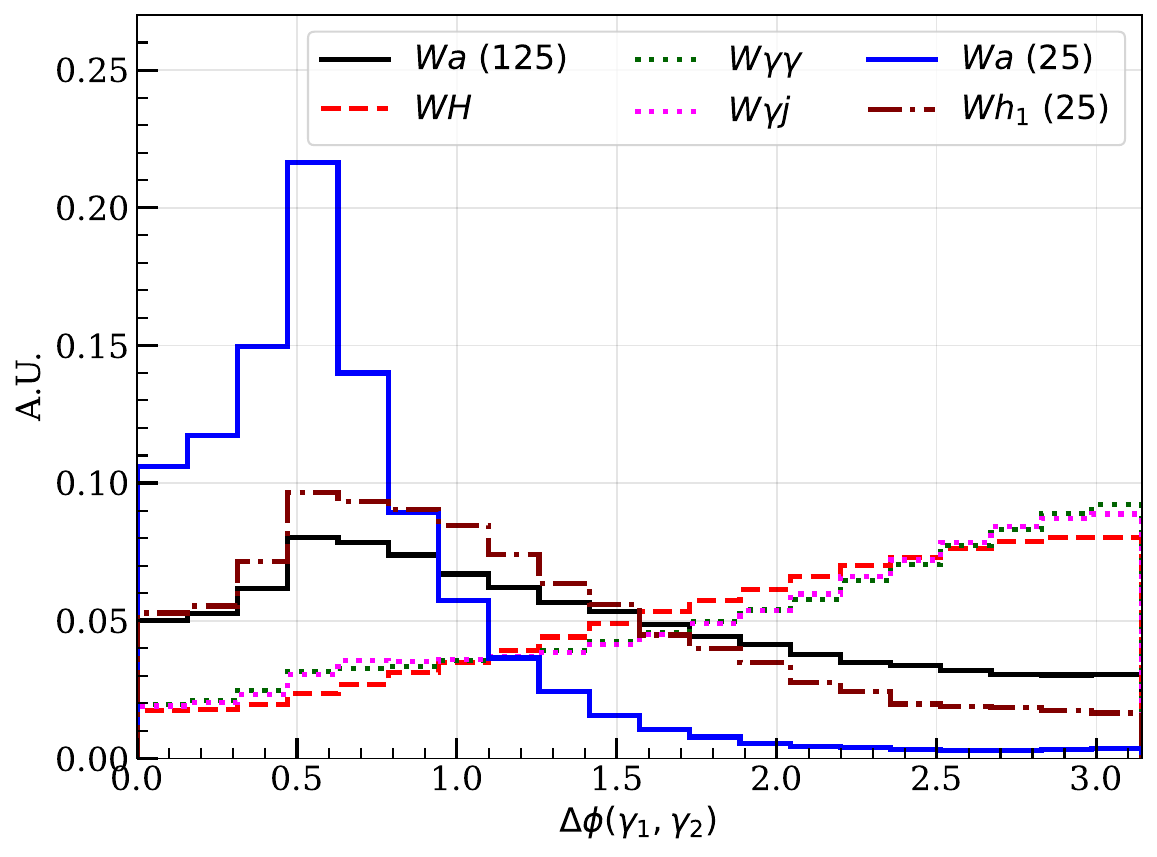}
		\vspace{1mm}
	\end{minipage}
	\caption{Normalized reconstruction-level distributions of the $W$-boson transverse
momentum, $p_T^W$ (left), and the azimuthal separation between the two
photons, $\Delta\phi(\gamma_1,\gamma_2)$ (right). The distributions are shown
for the $Wa$ signal with $m_a=25$ and $125~{\rm GeV}$, the mass-matched
NMSSM $Wh_1$ benchmark with $m_{h_1}=25~{\rm GeV}$, and the
$W\gamma\gamma$, $W\gamma j$, and $WH$ backgrounds. All samples are
subjected to the baseline selections in Eq.~\eqref{eq:gencuts}.}
	\label{fig:ptphi_reco}
\end{figure}

As discussed before, the derivative ALP field-strength interaction weights high-recoil configurations more strongly than a non-derivative Higgs-like interaction, and the corresponding SM backgrounds. Fig.~\ref{fig:ptphi_reco} shows the reconstructed 
$p_T^W$ distributions for the ALP scenario for two representative choices of the ALP masses, $m_a=25$ and $125$ GeV. As expected, the ALP events show a significantly enhanced high $p_T^W$ tail compared with the Higgs-like case and the other SM backgrounds. For a direct comparison at the same mass, we also show the reconstructed
$p_T^W$ distribution for a CP-even Higgs-like scalar with
$m_H=25~{\rm GeV}$ in the NMSSM benchmark discussed above.  Thus, $p_T^W$ provides information complementary to the helicity-angle observables, and directly probes the momentum dependence associated with the derivative structure of the ALP production vertex. It is also found to be very useful, both for distinguishing the ALP-like production from Higgs-like spin--0 production and for suppressing SM backgrounds.

We further consider azimuthal observables that provide complementary
sensitivity to the polarisation and helicity structure of the vector
boson. In this regard, we define $\Delta \phi_{\beta^\ast}$, the azimuthal
separation between the $V$ decay products in the $V\phi$ frame,
\begin{align}
\Delta \phi_{\beta^\ast} = | \phi_\ell^{V\phi}  - \phi_{\met/\ell}^{V\phi}|,
\label{eq:dphi_betastar}
\end{align}
which contains information about the boost and the production orientation of the vector boson.
Also, azimuthal correlations between the $V$ decay and scattering planes probe interference effects
among different helicity amplitudes \cite{ATLAS:2023zrv,Godbole:2011vw,ATLAS:2013xga,Bolognesi:2012mm}.
The related observables considered are $\phi_{\beta^\ast}$, the angle between the $V$ decay and scattering
planes in the same frame, and $\phi^{\ast}_{\ell V}$, the azimuthal
angle of the charged lepton in the $V$ rest frame with respect to the
production plane.

The azimuthal separation between the two photons, defined as
\begin{align}
\Delta\phi(\gamma_1,\gamma_2)
=
|\phi_{\gamma_1}-\phi_{\gamma_2}|,
\end{align}
is found to be very useful.
The corresponding distributions are shown in Fig.~\ref{fig:ptphi_reco}.  This
variable is correlated with the boost of the spin--0 resonance.  For a light
and boosted ALP, the two photons from $a\to\gamma\gamma$ become more
collimated, leading to smaller $\Delta\phi(\gamma_1,\gamma_2)$.  This feature
is particularly useful for suppressing continuum $V\gamma\gamma$ and
$V\gamma j$ backgrounds, where the photons do not generally originate from a
single boosted resonance.  Against resonant backgrounds such as
$VH(H\to\gamma\gamma)$, or against a mass-matched scalar look-alike, the
discriminating power of $\Delta\phi(\gamma_1,\gamma_2)$ is more benchmark
dependent and is mainly inherited from differences in the recoil spectrum of
the parent spin--0 state.

\section{Signal and Background Estimation}
\label{sec:analysis}
Armed with the observations from the previous sections, we now develop an
analysis strategy for the $W^\pm a$ and $Za$ signal channels and assess their
sensitivity at the HL-LHC.
We first define a cut-based signal region, which serves as a transparent and robust baseline for the analysis. Then further improve the signal sensitivity with a gradient-boosted decision-tree analysis implemented using the
\texttt{XGBoost} framework~\cite{Friedman_BDTG,Chen_2016}. The resulting signal sensitivities are subsequently translated into projected reaches on the electroweak ALP WCs.

The signal rate in the narrow-width approximation scales as
\begin{align}
\sigma(pp\to Va\to V\gamma\gamma)
\propto
\left(\frac{C_{\rm eff}}{f_a}\right)^2
{\rm BR}(a\to\gamma\gamma),
\label{eq:signal_scaling}
\end{align}
where $C_{\rm eff}=C_{WW}$ for the $W^\pm a$ channel and an electroweak
combination of $C_{WW}$ and $C_{BB}$ for the $Za$ channel.   

The analysis is divided into two mass regions, namely the low-mass region, 
$20\leq m_a\leq 80~{\rm GeV}$, and the high-mass region, $80< m_a\leq 200~{\rm GeV}$. 
This separation is motivated by the different kinematic properties(hardness) of the
final-state particles in the two mass regions.  We then impose the following
kinematic selections to define the signal regions and suppress the relevant
SM backgrounds.
 
\begin{itemize}
	\item 
	As discussed above, ALP production tends to populate harder recoil configurations, as illustrated in Fig.~\ref{fig:ptphi_reco}. We exploit this feature by imposing a harder requirement on the transverse momentum of the di-photon system, $p_T^{\gamma\gamma}$, which is approximately equivalent to the reconstructed $p_T^W$. For the lower ALP-mass region, we require $p_T^{\gamma\gamma}>50$ GeV, while a harder requirement of $p_T^{\gamma\gamma}>100$ GeV is imposed in the higher-mass region. This selection suppresses the SM backgrounds by a factor of approximately $\sim$ 2-3, while retaining a sizable fraction of the signal.
    \item 
     The azimuthal angle separation between two photons is also found to be one of the powerful variables to suppress the SM backgrounds further, which is also visible from the plot 
     Fig.~\ref{fig:ptphi_reco}. 
     \item 
     A requirement of $\met>30~{\rm GeV}$ is imposed in the $W^\pm a$ channel, where genuine missing transverse momentum arises from the neutrino in the leptonic $W$ decay. This requirement is particularly effective in suppressing the $Z\gamma\gamma$ background, which contains no genuine hard missing transverse momentum apart from detector and reconstruction effects.  However, no corresponding $\met$ requirement is imposed in the $Za$ channel, since it does not provide a significant improvement in the signal sensitivity.
     
     \item 
     We further impose requirements on polarization- and production-angle-sensitive variables, such as $\cos\theta^\ast_{\ell V}$ and $\cos\beta^\ast$.  These observables provide additional separation between ALP-like production, Higgs-like spin--0 production, and the SM backgrounds.
      \end{itemize} 

The cross sections(in fb) for the $W^\pm a$ signal and the corresponding backgrounds after successive selections are presented in Table~\ref{tab:ALP_yields_W}.
For the signal, the benchmark masses $m_a=20$ and $60~{\rm GeV}$ represent the low-mass region, while $m_a=100$, $125$, and $200~{\rm GeV}$ represent the high-mass region.
The first row of the table presents the raw production cross sections for the signal processes without applying any selection cuts, along with those for the background processes after applying the initial cuts, as specified in the figure caption.
The second row includes the same after multiplying by the relevant decay branching ratios.  The subsequent rows show the cross sections after applying each of the successive selection requirements listed in the first column.
Notice that the cut on $p_T^{\gamma\gamma}$ is very effective in suppressing the SM backgrounds while retaining a sizable fraction of the signal cross sections. 
The azimuthal selection provides substantial suppression of continuum backgrounds.
The subsequent cuts on $\cos\theta^\ast_{\ell\gamma\gamma}$ and $\cos\beta^\ast$, which exploit the polarization and production-angle structures discussed in Sec.~\ref{sec:simulation_observables}, have only a mild effect on the overall yields. Their primary role is to discriminate between the different signal hypotheses.
To present the signal sensitivity 
for each mass point, assuming an integrated luminosity option, 
${\cal L}$=3 ab$^{-1}$,
we present the significance calculated as \cite{Cowan:2010js}, 
\begin{equation}
\mathcal{Z} =\sqrt{2\left[
(S+B)\ln\left(\frac{(S+B)(B+\sigma_B^2)}{B^2+(S+B)\sigma_B^2}\right)
-\frac{B^2}{\sigma_B^2}\ln\left(1+\frac{\sigma_B^2 S}{B(B+\sigma_B^2)}\right)
\right]},
\label{eq:significance}
\end{equation}
where $S$ and $B$ denote the number of signal and background events, respectively, for a given luminosity, and $\sigma_{B}$ is the systematic uncertainty in background estimation, which is considered $0\%$ and $10\%$ of B.
After all selections, the signal significances are presented assuming background uncertainties$(\sigma_b)$ corresponding to our conservative choices of benchmarks, Eq.\eqref{eq:alp_benchmark}.
For $\sigma_b = 0\%$, the $W^\pm a$ channel reaches significances ($\mathcal{Z}$) $\sim 4$ for the range $m_a=80$--$125~{\rm GeV}$. The significance decreases to $\mathcal{Z}\simeq 2.2$ for
$m_a=200~{\rm GeV}$, mainly because of the smaller production rate and diphoton branching fraction.  At lower masses, the sensitivity is also
reduced, with $\mathcal{Z}\simeq 1.2$ and $3.6$ for
$m_a=20$ and $60~{\rm GeV}$, respectively.
The reduced sensitivity at the lowest masses is mainly due to the softer
photons from the ALP decay, which lead to a lower acceptance under the
baseline photon transverse-momentum requirements. 
For $\sigma_b =10\%$, the significance reduces drastically. 
However, for optimistic choices of benchmark values of $C_{WW}$ and
$C_{BB}$, the signal significances are found to be high, as shown 
in the last row of the table. 
\begin{table}[htb!]
	\centering
	\resizebox{\textwidth}{!}{%
		\begin{tabular}{|l|ccccccccc|} 
			\hline
			\textbf{Cut} & \textbf{20} & \textbf{60} & \textbf{100}  &  \textbf{125} &\textbf{200}&
			\textbf{$HW$}& \textbf{$W\gamma\gamma$} & \textbf{ $W\gamma j$} & \textbf{$Z\gamma \gamma$}\\
			\hline
			CS                    & 5.9  & 3.48 &  2.25   &  1.76  & 0.95 & 1222 & 130 & $7.6 \times 10^{4}$ & 170.5   \\
			CS $\times$ BR        & 1.23  & 0.7 &  0.46   &  0.35  & 0.16 & 0.4 & 33.1 & $2 \times 10^{4}$ & 11.4   \\
			$\ell\gamma\gamma$    & 0.069  & 0.144 &  0.13   &  0.108  & 0.06         & 0.10 & 4.09 & 5.44 & 1.86  \\
			$p_T^{\gamma\gamma}>$ 50 GeV   & 0.047 & 0.124 & 0.091    &  0.083  & 0.051 & 0.08 & 1.49 & 3.29 & 0.66 \\
			~~~~~$>$ (100 GeV)& & & &  & &                                             (0.041) &(0.52) &(1.26) &(0.158) \\
		 $0 <\Delta\phi(\gamma_1,\gamma_2) < 2$ & 0.047 & 0.118 & 0.085    & 0.075 &0.04 & 0.047  & 0.83 & 1.83 & 0.48 \\
			 & & &   & & &                                                                (0.035) &(0.26) &(0.69) &(0.11) \\
			$\slashed{E}_T >$ 30 GeV   & 0.031 & 0.089 & 0.069    &0.062 &  0.035     & 0.038 & 0.48  &1.19 & 0.05  \\
				&  & & & & &                                                           (0.03) &(0.17) &(0.5) &(0.018) \\
			$-0.1 < \cos\theta^\ast_1 < 0.8$ & 0.031 & 0.087 & 0.068 &  0.06 &0.033   &0.038 & 0.45 & 1.15 & 0.05\\
			 & & &&  & &                                                                 (0.03) &(0.16) &(0.48) &(0.017) \\
			$-0.8 < \cos\beta^\ast < 0.8$  & 0.026 & 0.075 & 0.061     &0.055 & 0.03    &0.035 & 0.25 & 0.92 & 0.04  \\
			 & & & && &                                                                  (0.028) &(0.1) &(0.42) &(0.013) \\
			\hline
			$\mathcal{Z}$ for $\sigma_{b}=0\%$   & 1.2 & 3.6 & 4.4 & 4.0 & 2.2       & & & &\\
			$\mathcal{Z}$ for $\sigma_{b}=10\%$  & 0.2 & 0.6 & 1.0 & 1.0 & 0.5     & & & &\\
			\hline
            $\mathcal{Z}|_{\sigma_{b}=10\%}$ with $C_{WW}= 0.09, C_{BB} =0.3$ & 3.6 & 9.0 & 14 & 13 & 8  & & & &\\
            \hline
		\end{tabular}
	}
	\caption{\small Cross sections(CS) in fb for the $W^\pm a$ analysis at $\sqrt{s}=14~{\rm TeV}$ for a set of 
$m_a=20$, $60$, $100$, $125$, and $200~{\rm GeV}$, corresponding to 
the the ALP benchmark, Eq.~\eqref{eq:alp_benchmark}. The first row gives the raw production cross sections.  The continuum $W\gamma\gamma$, $W\gamma j$, and
$Z\gamma\gamma$ processes are generated with
$p_T^\gamma>10~{\rm GeV}$, $p_T^j>20~{\rm GeV}$,
$|\eta_{\gamma,j}|<2.5$, and
$\Delta R(\gamma/j,\gamma)>0.3$. The CS presented in the second row includes the relevant decay
branching fractions. Subsequent rows show the CS after the
successive selections listed in the first column, in addition to the baseline
requirements of Eq.~\eqref{eq:gencuts}. The entries in parentheses correspond
to the background yields obtained with the
$p_T^{\gamma\gamma}>100~{\rm GeV}$ requirement used for the high-mass
region.The expected significances($\mathcal{Z}$) are evaluated
for ${\cal L}=3~{\rm ab}^{-1}$ using Eq.~\eqref{eq:significance}, assuming
background systematic uncertainties of $0\%$ and $10\%$. The final row shows
the corresponding $10\%$-systematic significance for the alternative
benchmark $(C_{WW}, C_{BB})=(0.09,0.3)$, assuming the same 
signal acceptance efficiencies corresponding to the 
benchmark values of coupling strengths (Eq.~\eqref{eq:alp_benchmark}.
} 
	\label{tab:ALP_yields_W}
\end{table}
The corresponding $Za$ results are shown in Table~\ref{tab:ALP_yields_Z}. The same procedure is followed, 
with $p_T^{\gamma\gamma}>100~{\rm GeV}$ for the higher-mass benchmarks 
and $p_T^{\gamma\gamma}>50~{\rm GeV}$ for the lower-mass benchmarks. The 
$Z$ channel benefits from a fully reconstructible dilepton system and a
cleaner event topology, but its rate is significantly lower than that 
of the charged-current channel.  Consequently, the final cut-based 
significances are lower. For instance, for the higher-mass benchmarks they are in the range 
$\sim 2-1$, for $m_a=80-200~{\rm GeV}$, while in the lower-mass region 
they remain below the $2\sigma$ level.  
The irreducible $Z\gamma\gamma$ background remains the dominant background 
even after selection, whereas the reducible $Z\gamma j$ contribution 
is efficiently suppressed by the jet-to-photon fake rate. In this case too, the signal significances improve in the low-mass region for the optimistic benchmark choices.

\begin{table}[htb!]
	\centering
	\resizebox{\textwidth}{!}{%
		\begin{tabular}{|l|cccccccc|}
			\hline
			\textbf{Cut} & \textbf{20}  & \textbf{60} & \textbf{100}  &  \textbf{125} &\textbf{200} &
			\textbf{$HZ$} & \textbf{$Z\gamma\gamma$} & \textbf{ $Z\gamma j$}\\
			\hline
CS                                  & 3.36 &  2.06          &  1.36  & 1.067    & 0.58        & 642.4   & 170.5   &  $2.7 \times 10^{4}$\\            
CS $\times$ BR                      & 0.22 &  0.13          &  0.086  & 0.067    & 0.03        & 0.064   & 11.4   &  $2.1 \times 10^{3}$\\
$\ell\ell\gamma\gamma$              & 0.008 &  0.018        &  0.016 & 0.014   & 0.007       & 0.012   & 1.751   &  0.16 \\ 
$p_T^{\gamma\gamma}>$ 50 GeV        & 0.006 &  0.016        &  0.012 &  0.011  & 0.006       & 0.009  &  0.609 &  0.093\\
~~~~~$>$ (100 GeV) 	& & & &  & &                                                              (0.005) &  (0.179) & (0.033)    \\
$0 <\Delta\phi(\gamma_1,\gamma_2)<2$& 0.006 &  0.015        &  0.011 &  0.01  & 0.005       & 0.006  &  0.451  &  0.066  \\
	            & & & &  & &                                                                   (0.004) &  (0.133) & (0.025)\\
$-0.1 < \cos\theta^\ast_1 < 0.8$    &0.005  &  0.014        &  0.01  &  0.009 & 0.005       & 0.005  &  0.413 &  0.061\\
                & & & &  & &                                                                   (0.004)&   (0.123)& (0.023)\\

$-0.8 < \cos\beta^\ast < 0.8$      &0.004  &   0.012        &  0.009  & 0.008  &  0.004      & 0.005  & 0.34  & 0.051\\
			 & & & & & &                                                                         (0.004)  & (0.106) &(0.02) \\
\hline
$\mathcal{Z}$ for $\sigma_{b}=0\%$     &  0.4   & 1.0   & 1.4   &  1.2  & 0.6        & & &\\
$\mathcal{Z}$ for $\sigma_{b}=10\%$    &  0.1   & 0.3   & 0.6   &  0.6   & 0.3        & & &\\
\hline
$\mathcal{Z}|_{\sigma_{b}=10\%}$ with $C_{WW}= 0.09, C_{BB} =0.3$ & 2.6 & 6.4 & 12 & 11 & 6.1   & & &\\
\hline
\end{tabular}
}
\caption{ Same as Table~\ref{tab:ALP_yields_W}, but for the $Za$ channel.}
\label{tab:ALP_yields_Z}
\end{table}

As discussed above, we performed a cut-based analysis to study how the
different selection requirements, including polarization-sensitive angular
variables, suppress the backgrounds and improve the signal significance.
Several features are common to both channels.  First, the di-photon recoil cut 
captures the harder spectrum induced by the derivative ALP interaction and 
is therefore particularly effective. Second, the polarization- and angular-motivated 
cuts improve the signal-to-background ratio, but they do not always maximize 
the statistical significance when applied as independent rectangular cuts, 
because they also remove part of the signal acceptance. Third, the 
$W^\pm a$ channel provides the stronger discovery reach because of its larger 
production rate, whereas the $Za$ channel remains valuable as an independent 
and cleaner probe of the electroweak ALP coupling combinations. 
Notably, we do not use the di-photon invariant mass,
$m_{\gamma\gamma}$, as a signal-selection variable.
Experimentally, an ALP signal is expected to appear as a resonant structure in the
$m_{\gamma\gamma}$ distribution over the continuum background. Selecting
events within an appropriate invariant-mass window around the resonance could
therefore substantially improve the signal significance.  However, we do not
impose such a requirement in the event selection, since the ALP mass is not
known \emph{a priori}. Applying such a requirement would bias the analysis toward a specific ALP mass hypothesis and thereby reduce the model independence of our search strategy. Our initial goal is to explore the feasibility of detecting the ALP  signal over a given mass range at the HL-LHC. 

The cut-based analysis discussed above motivates the use of a multivariate
analysis (MVA) to further improve the sensitivity to the ALP signal. We employ
XGBoost (Extreme Gradient Boosting)~\cite{Chen_2016}, a widely used
machine-learning algorithm based on an ensemble of decision trees
\cite{Friedman_BDTG}. XGBoost can efficiently capture nonlinear correlations
and complex patterns among the input variables, making it well suited for
signal--background discrimination in collider analyses. In this MVA, the
kinematic, recoil, and polarization-sensitive observables discussed above are
combined with additional relevant kinematic features, including
\begin{equation}
\begin{aligned}
\textbf{Observables:}\quad&
p_{T}^W,
\Delta \phi (\gamma_1, \gamma_2),
\met,
\Delta \phi (\gamma\gamma,  W),
M_T(\ell,\met) \equiv \sqrt{2p_T^\ell \met(1-\cos\Delta \phi(\ell , \met))},\\
&
\Delta \eta(\ell,\met),
\Delta \phi (\ell , \gamma \gamma),
\Delta \phi (\met, \gamma \gamma), \\
&
m_{\gamma \gamma},
\Delta \phi(\ell , \met),
\Delta \eta(\gamma \gamma , W),
p_{T \ell},
p_{T \gamma\gamma},\\
&
\cos\theta_{\ell\gamma\gamma}^\ast,
\cos\theta_{\ell V}^\ast,
\cos\beta^\ast,
L_P,
\Delta\phi_{\beta^\ast},
\phi_{\beta^\ast},
\phi^{\ast}_{\ell V},
\\
\end{aligned}
\label{eq:observables}
\end{equation}

and are used as inputs for classification after standardization, with one variable removed from each pair exhibiting a correlation greater than $70\%$.
For both signal and background events, about 70\% of events are used for 
training, whereas the rest of the events are used for testing.
Each event is assigned a weight proportional to
$\sigma_i{\cal L}/N_i$, where $\sigma_i$ and $N_i$ denote the cross section
and the number of generated events for sample $i$, respectively. The
background samples are then combined into a single background class using
these event weights. Both signal and background events are subjected to the
baseline selections in Eq.~\eqref{eq:gencuts}. After these selections,
approximately $(1$--$2)\times10^{5}$ signal events and
$(3$--$5)\times10^{5}$ background events per process remain available for
the analysis.
The XGBoost is trained separately for
two cases:
\begin{description}
    \item[\textbf{Case A}] a baseline training in which $m_{\gamma\gamma}$ is
    not included as an input feature;
    \item[\textbf{Case B}] an extended training in which $m_{\gamma\gamma}$ is
    included as an input feature.
\end{description}
During training, the classification error and AUC are monitored, with early stopping after 50 consecutive boosting rounds without improvement 
to reduce the risk of overtraining.
\begin{figure}[t]
	\centering
	\begin{minipage}{0.48\textwidth}
		\centering
		  \includegraphics[width=\linewidth]{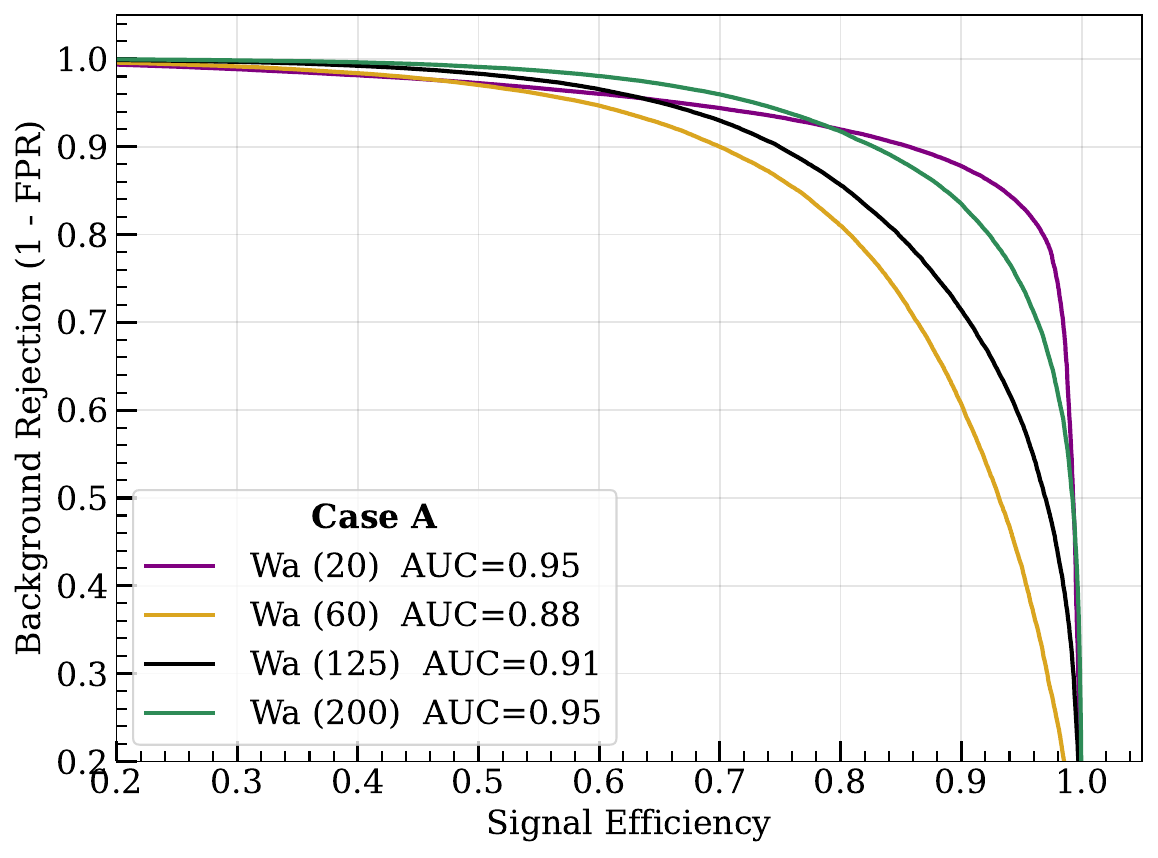}
		\vspace{1mm}
	\end{minipage}
	\hfill
	\begin{minipage}{0.48\textwidth}
		\centering
		  \includegraphics[width=\linewidth]{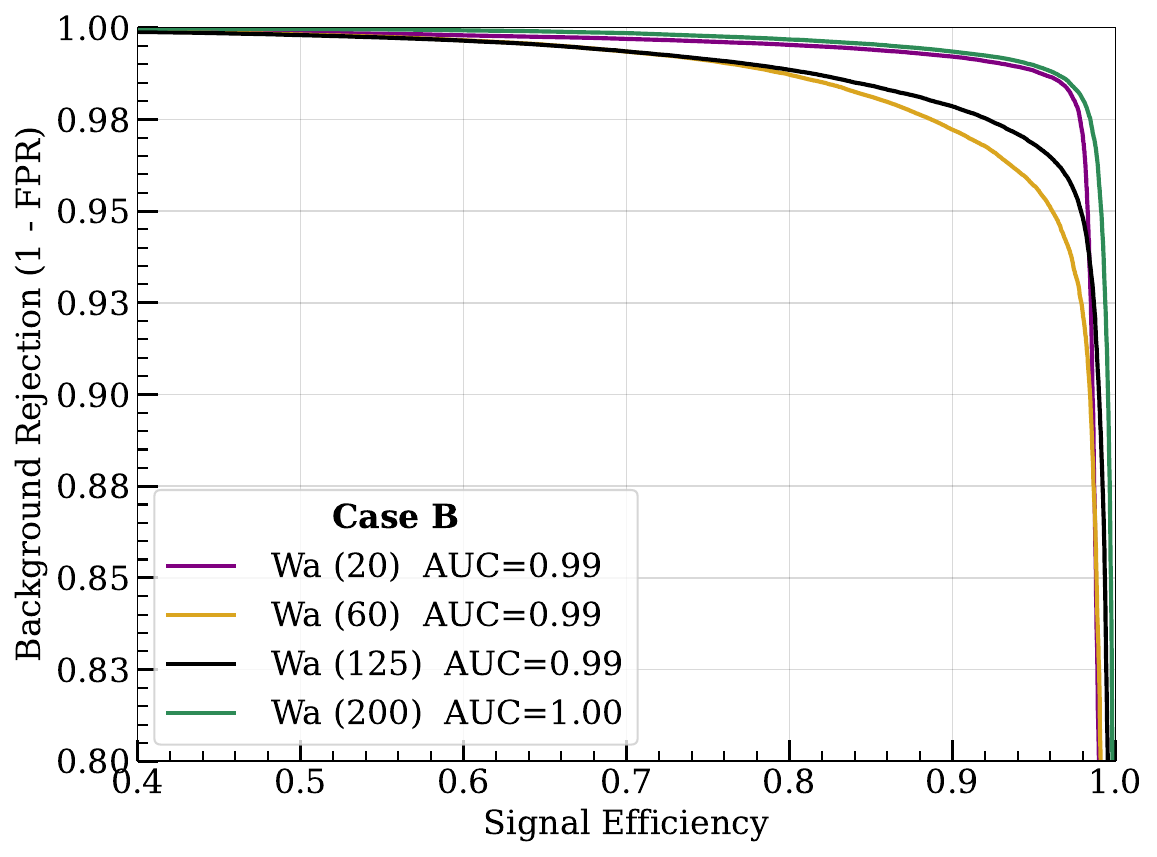}
		\vspace{1mm}
	\end{minipage}
	\caption{Receiver operating characteristic (ROC) curves for the $W^\pm a$ analysis
at the benchmark masses $m_a=20$, $60$, $125$, and $200~{\rm GeV}$. A
separate XGBoost classifier is trained at each mass point to distinguish the
$W^\pm a$ signal from the combined SM backgrounds. \textit{Left}: Case A,
with $m_{\gamma\gamma}$ excluded from the input features. \textit{Right}:
Case B, with $m_{\gamma\gamma}$ included as an additional input feature.
The same XGBoost hyperparameters are used for all benchmark masses and in
both cases.}
	    \label{fig:rocs_W}
\end{figure}

\begin{figure}[t]
	\centering
	\begin{minipage}{0.48\textwidth}
		\centering
		  \includegraphics[width=\linewidth]{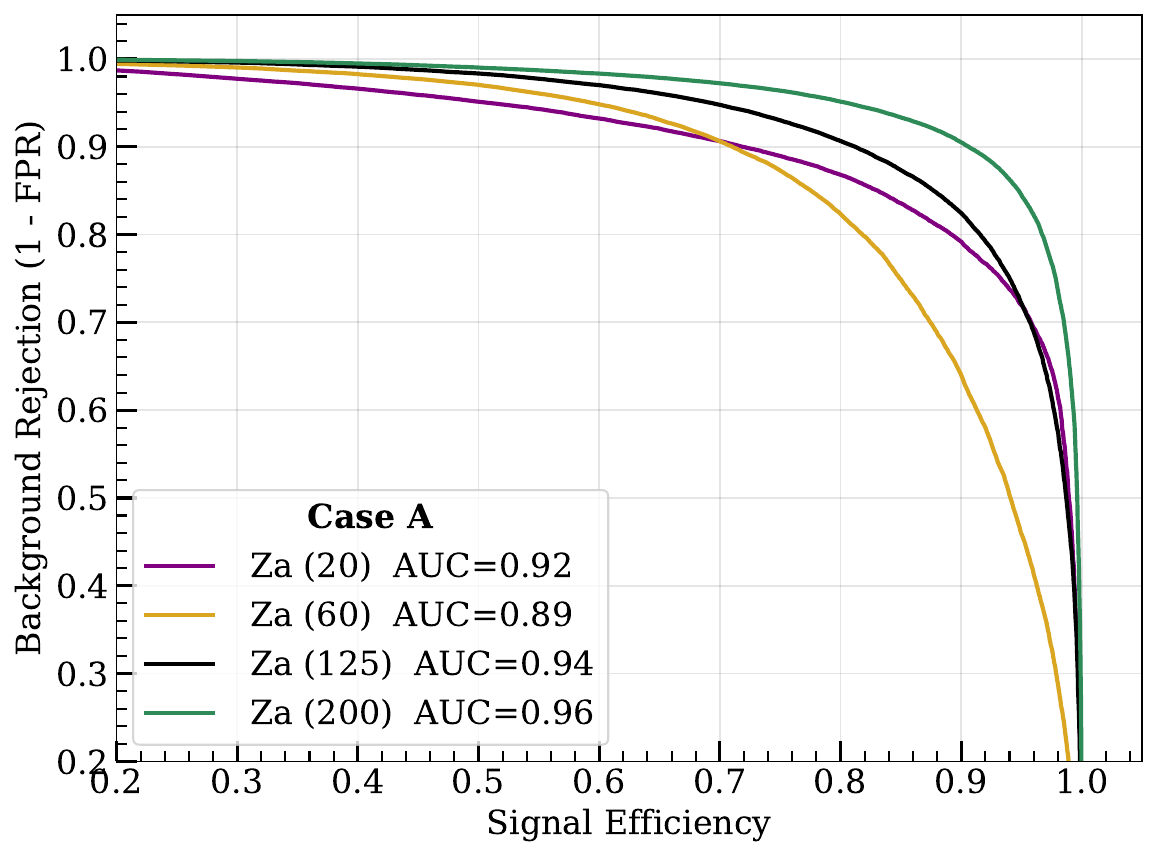}
		\vspace{1mm}
	\end{minipage}
	\hfill
	\begin{minipage}{0.48\textwidth}
		\centering
		  \includegraphics[width=\linewidth]{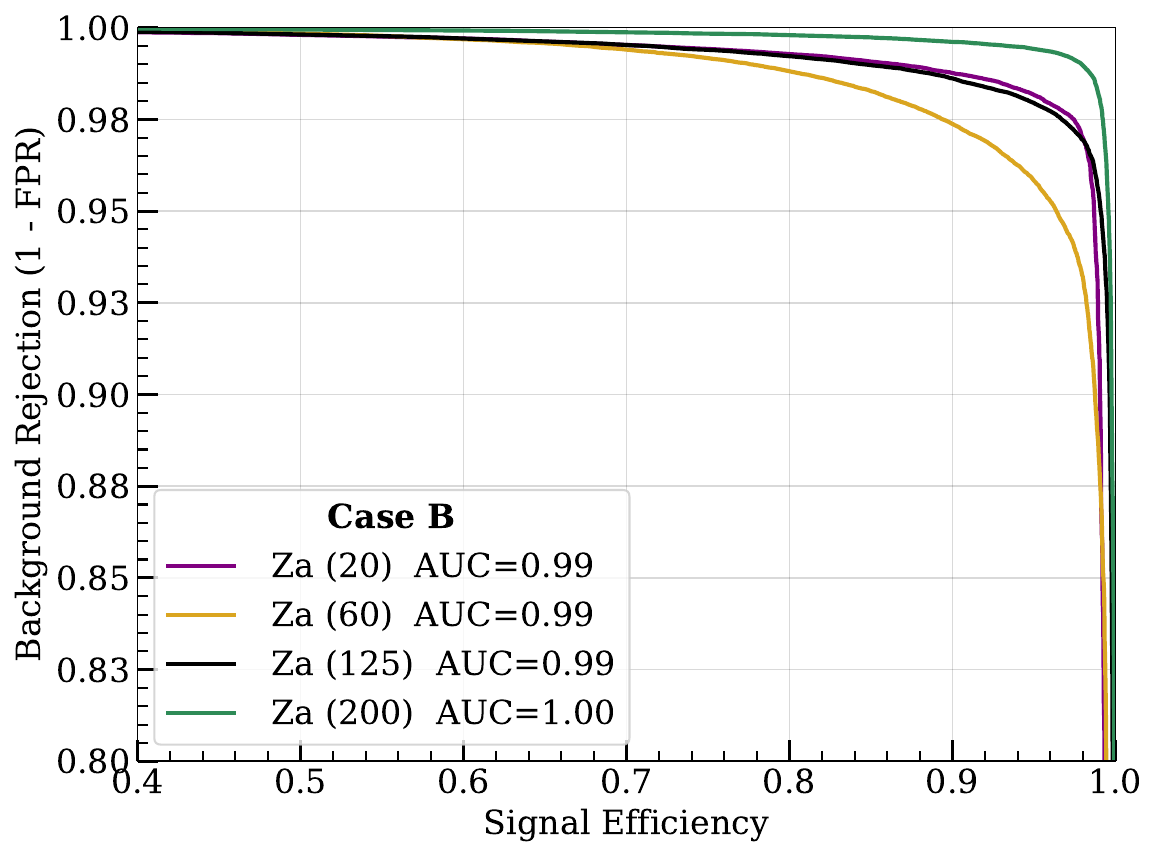}
		\vspace{1mm}
	\end{minipage}
	\caption{ROC curves for the $Za$ analysis.
Case A and Case B are defined as in the $W^\pm a$ analysis shown in
Fig.~\ref{fig:rocs_W}.}
	\label{fig:roc_Z}
\end{figure}

The resulting ROC curves for the $W^\pm a$ channel are shown in
Fig.~\ref{fig:rocs_W}. The left panel corresponds to Case A, in which
$m_{\gamma\gamma}$ is excluded from the input feature set. This setup tests
the discriminating power of the non-resonant event information without
relying directly on the diphoton mass peak. The classifier achieves good
signal--background separation over most of the considered mass range, with
AUC values in the range $0.88$--$0.95$ for the benchmark masses shown.
This demonstrates that the event kinematics, including the recoil and
polarization-sensitive observables, carry substantial discriminating power
independently of the resonant diphoton-mass information. A similar behaviour
is observed for the $Za$ channel, as shown in Fig.~\ref{fig:roc_Z}.
The right panel shows the corresponding training after including
$m_{\gamma\gamma}$ as an additional input feature.  As expected, the
performance improves substantially, reaching AUC values close to $1$. 
This improvement is largely driven by the ability of the classifier to exploit the resonant di-photon mass peak of the signal over the continuum backgrounds. Such a mass-assisted training provides an estimate of
the maximal achievable discrimination for a fixed mass hypothesis.

\begin{figure}[htb!]
 	\centering
 	\includegraphics[width=0.6\linewidth]{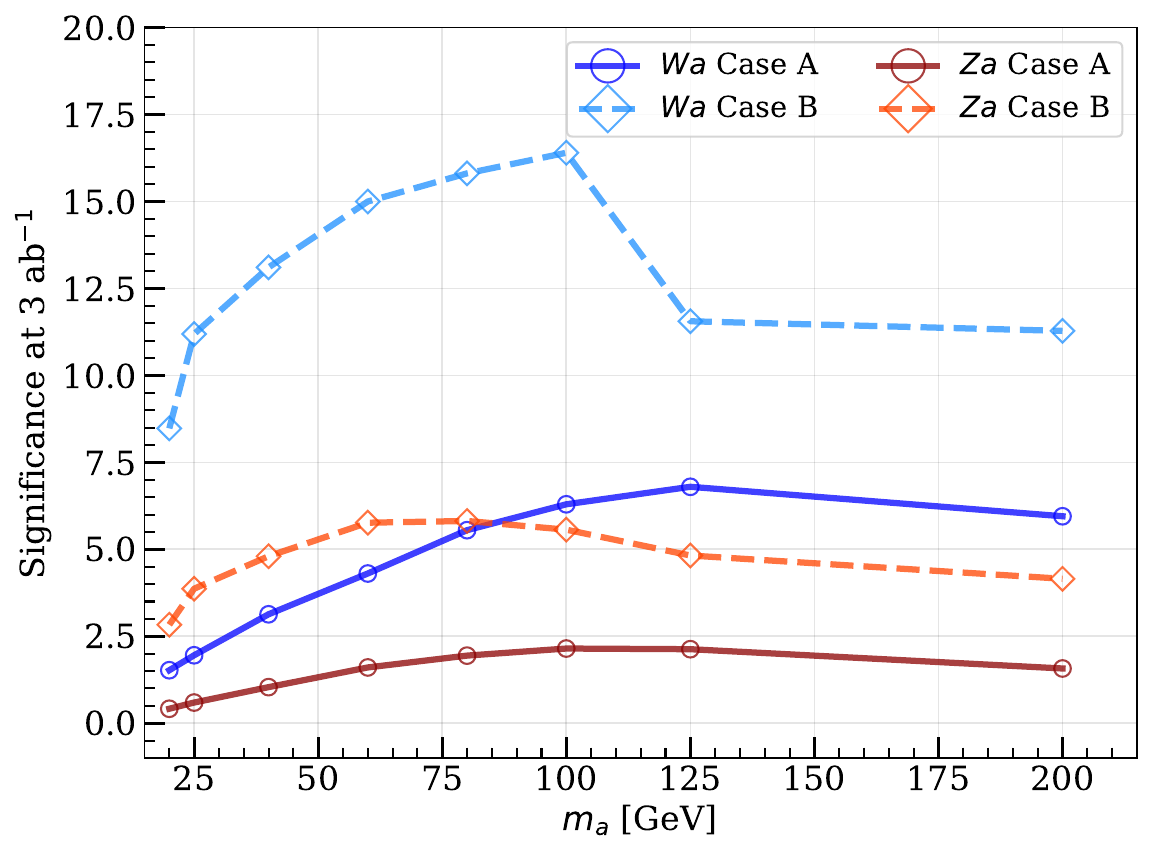} 
 	\caption{ Expected significance as a function of $m_a$ for the $W^\pm a$ and $Za$
channels in Case A and Case B, obtained from the XGBoost analysis for
${\cal L}=3~{\rm ab}^{-1}$ and assuming a $10\%$ systematic uncertainty
on the total background.}
 	\label{fig:signifi}
 \end{figure}

Based on the XGBoost analysis, we evaluate the signal significance for a range of masses from 20 - 200 GeV for an integrated luminosity ${\cal L}=3~\mathrm{ab}^{-1}$ option, and it is presented in 
Fig.~\ref{fig:signifi} for both Wa and Za channels. 

In calculating the statistical significance, we include an overall 10\% uncertainty in the background estimation. At $m_a=125~{\rm GeV}$, the significance in Case B is reduced because the SM Higgs background, $H\to\gamma\gamma$, produces a resonant peak in the same diphoton-mass region, thereby reducing the discriminating power gained from $m_{\gamma\gamma}$. At higher ALP masses, the sensitivity decreases primarily because of the smaller signal production cross section, whereas at lower masses the final-state leptons and photons tend to be softer, leading to a reduced selection efficiency. A similar trend is observed in the $Za$ channel, although with a lower overall
sensitivity. Including $m_{\gamma\gamma}$ in Case B significantly improves the
significance, reaching about $5$--$6$ for $m_a\simeq60$--$100~{\rm GeV}$,
before decreasing at larger masses because of the reduced signal rate.
 
\subsection*{ALP Coupling Sensitivity}

Having obtained sizable signal sensitivity over the considered mass range, we now translate the expected signal yields into projected sensitivities to the ALP electroweak couplings. For each ALP mass hypothesis, the XGBoost classifier provides the expected signal and background yields, denoted by $S_0$ and $B_0$, respectively, for the benchmark parameters given in Eq.~\eqref{eq:alp_benchmark}. Since the background yield is independent of the ALP coupling parameters, it is
taken to be the same for all choices of couplings.

For the $Wa$ channel, since the associated-production amplitude is linear in $C_{WW}$, the production cross section scales quadratically with the coupling. The signal yield at a given coupling value is therefore obtained by rescaling the benchmark yield as,
\begin{equation}
    S(k) = S_0 \left( \frac{k}{k_0} \right)^2,
    \qquad {\rm with}\quad  
    k \equiv C_{WW}, \qquad
    k_0 \equiv C_{WW}^{0}.
    \label{eq:signal_yield_scaling}
\end{equation}
Here, the signal selection efficiency and the relevant branching ratio are assumed to remain unchanged under the rescaling. At higher masses, the opening of additional electroweak decay channels can introduce a non-trivial dependence of the diphoton branching ratio on $C_{WW}$ and $C_{BB}$; therefore, a full reinterpretation beyond this fixed-branching-ratio rescaling would require accounting for this dependence.

\begin{figure}[t]
	\centering
	\begin{minipage}{0.48\textwidth}
		\centering
		  \includegraphics[width=\linewidth]{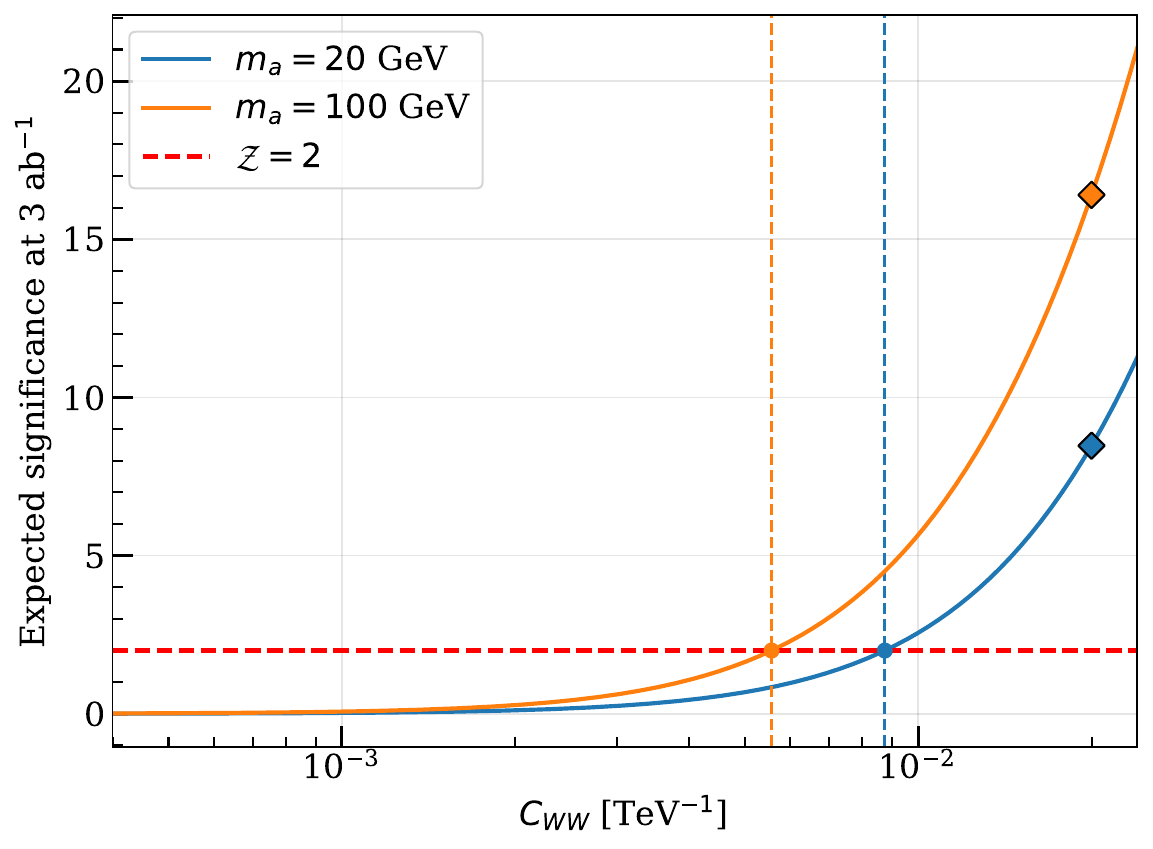}
		\vspace{1mm}
	\end{minipage}
	\hfill
	\begin{minipage}{0.48\textwidth}
		\centering
		  \includegraphics[width=\linewidth]{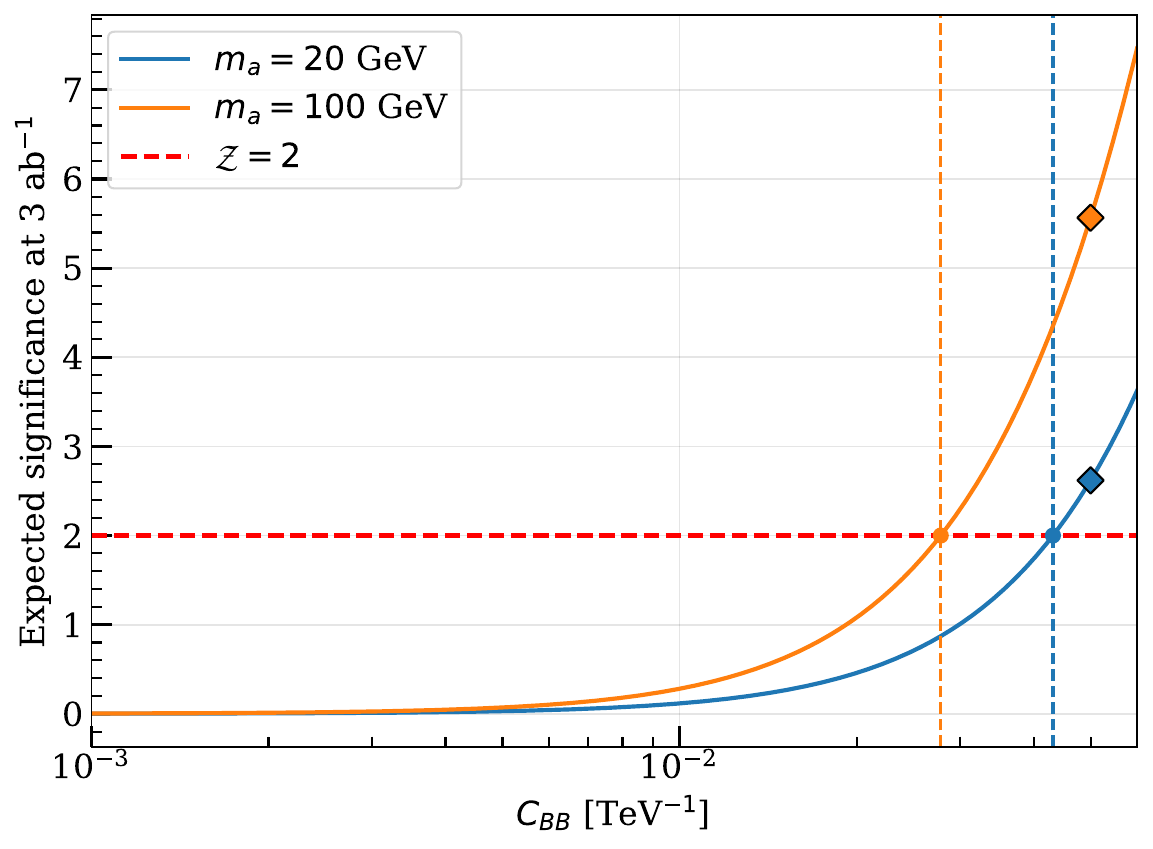}
		\vspace{1mm}
	\end{minipage}
	\caption{Expected Asimov significance as a function of the effective couplings $C_{WW}$ (left) and $C_{BB}$ (right) for $m_a=20$ and $100~{\rm GeV}$. The signal yields are rescaled according to Eq.~\eqref{eq:signal_yield_scaling}, while the backgrounds are kept fixed. The horizontal dashed line denotes $Z=2$, corresponding to the expected 95\% C.L. sensitivity, while the vertical dashed lines indicate the corresponding coupling reaches. The markers denote the benchmark points used to generate the signal samples.}
	\label{fig:coupling_reach}
\end{figure}

The expected sensitivity is evaluated using the Asimov profile-likelihood significance Eq.~\eqref{eq:significance} with $\sigma_B = B \times10\% $. The expected 95\% confidence level sensitivity is then defined by the coupling value satisfying $\mathcal{Z}(k_{95}) =2$. This equation is solved numerically, and the significance is evaluated over a logarithmically spaced scan of the coupling parameter $k$. The intersection of the resulting $\mathcal{Z}(k)$ curve with $\mathcal{Z}(k_{95})$ provides the best expected minimum value of the coupling strength that can be measured Fig.~\ref{fig:coupling_reach} (left). For the $Wa$ channel at $m_a=100~{\rm GeV}$, the benchmark yields are
$S_0=184$ and $B_0=32$ for Case B. Using these yields, we obtain a projected $95\%$ CL sensitivity of $C_{WW}\gtrsim 5.57\times10^{-3}~{\rm TeV}^{-1}$. The resulting coupling sensitivity varies only moderately over the 20-200 GeV mass range, reflecting the relatively mild variation of the
signal significance between the different benchmark masses.

 In contrast to the $Wa$ channel, the neutral electroweak interactions of the ALP involve both $\sigma (pp \to Za) \propto |\mathcal{M}(C_{WW}, C_{BB})|^2$ since the process is mediated by both Z and $\gamma$. Hence, instead of an independent measure of both the couplings, we choose a one-dimensional benchmark direction in the $(C_{WW}, C_{BB})$ parameter space as,
\begin{equation}
    (C_{WW},C_{BB})
    =\lambda\,(C_{WW}^{0},C_{BB}^{0})
    =\lambda\,(0.02,0.05) ~\mathrm{TeV}^{-1},
    \label{eq:benchmark_ray}
\end{equation}
where $\lambda=1$ corresponds to the benchmark point used in the simulation. Varying $\lambda$ therefore changes the overall strength
of the ALP electroweak interactions. 
The signal yield for a given value of $\lambda$
can be scaled from $S_{\mathcal{Z},0}$, i.e. the signal yield at $\lambda=1$, as,
\begin{equation}
    S_\mathcal{Z}(\lambda)=S_{\mathcal{Z},0}\lambda^2.
    \label{eq:Za_signal_scaling}
\end{equation}
The limits on the WCs are similarly obtained by solving $\mathcal{Z}=2$. Fig.~\ref{fig:coupling_reach} (right) shows the limit on $C_{BB}$ for two mass points, 20 and 100 GeV. For the same $m_a = 100$ GeV with $S_0 = 30$ and $B_0 = 15$, the expected $95\%$ CL limit on $C_{BB} \gtrsim 2.78 \times 10^{-2} ~ \mathrm{TeV}^{-1}$ and $C_{WW} \gtrsim 1.11 \times 10^{-2} ~ \mathrm{TeV}^{-1}$ . For the projected coupling reach, we adopt the Case~B results.  Relative to the current constraint, this leads to an order-of-magnitude improvement in the sensitivity to $C_{WW}$ and $C_{BB}$~\cite{Carra:2021ycg, Gavela:2019cmq}.   
\section{Model Discrimination}
\label{sec:model_disc}
The analysis discussed above shows that the
reasonable ALP signal sensitivity for the considered benchmarks can be achieved at the
HL-LHC in the $V\gamma\gamma$ final state. More precisely, the ALP signal is
expected to be visible as a distinct resonance peak over the continuum SM
background. However, an excess in the $V\gamma\gamma$ final state, observed
only as a peak in the diphoton invariant-mass distribution, would not be
sufficient to determine the nature of the underlying spin--0 resonance. An identical
signature could also arise from the associated production of a light scalar with a comparable mass, as occurs in several BSM models with extended Higgs sectors.
We therefore develop a strategy to discriminate between these signal
hypotheses. For the sake of presentation, we consider an extended Higgs sector containing two Higgs doublets and an additional singlet, commonly referred to as a 2HDM+S. Such a scalar sector can accommodate additional light neutral Higgs
states alongside the observed SM-like Higgs boson. A well-motivated
supersymmetric realization of this structure is provided by the
Next-to-Minimal Supersymmetric Standard Model
(NMSSM)~\cite{Ellwanger:2009dp,Guchait:2015owa}. Depending on the model
parameters, the additional neutral states can be substantially lighter than
the SM-like Higgs boson, and their associated production followed by decay
into two photons can mimic the ALP signature considered here ~\cite{Guchait:2015owa,Guchait:2016pes}. Establishing
the origin of an observed signal therefore requires investigating whether the
event kinematics are consistent with the predictions of a given model. In
particular, discriminating between the different Lorentz structures of the
couplings involved in the production and decay processes is crucial for
identifying the nature of the underlying resonance, rather than merely
establishing its presence.
 
We use the NMSSM as a phenomenological Higgs-like scalar benchmark to simulate the $V\gamma\gamma$ signal in Eq.~\eqref{eq:sim_processes}. The coupling of
the light scalar $h_1$ to electroweak gauge bosons has the SM Higgs form, rescaled by a factor $\kappa_V$, 
\begin{align}
g_{h_1 VV} &= \kappa_V\,\frac{2m_V^2}{v};\quad V=W,Z
\end{align}
The factor $\kappa_V$ is model dependent,
and controls the overall production rate. 
Clearly, this non-derivative coupling is qualitatively different from the
dimension-five, momentum-dependent ALP interaction
$aV_{\mu\nu}\widetilde V^{\mu\nu}$, Eq.~\eqref{eq:aVV_vertex} which preferentially produces
transversely polarized associated gauge bosons, as discussed above.

\begin{figure}[htp!]
        \centering
\includegraphics[width=0.6\linewidth]{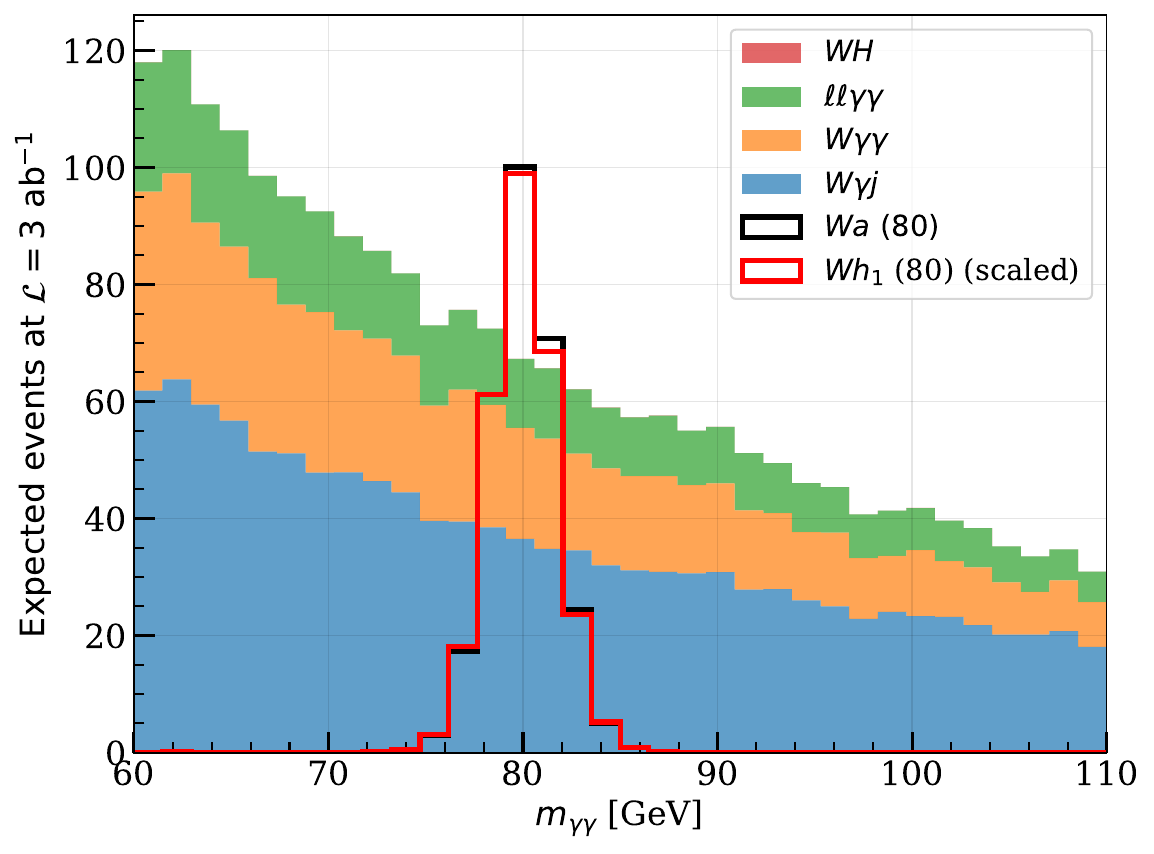} 
 \caption{Reconstructed diphoton invariant-mass distribution 
 for the $Wa$ and $Wh_1$ hypotheses with $m_a=m_{h_1}=80~{\rm GeV}$ along with the SM backgrounds. These are subject to the baseline selections of
Eq.~\eqref{eq:gencuts} and $\Delta\phi(\gamma_1,\gamma_2)<1.5$.
}
        \label{fig:disvar}
\end{figure}

To enable a direct comparison between the two signal hypotheses, we choose the resonance masses to be identical and rescale $\kappa_V$ such that the visible $V\gamma\gamma$ production rate matches that of the ALP benchmark, i.e 
\begin{align}
m_{h_1}=m_a,
\qquad
\sigma(pp\to Vh_1)\,{\rm BR}(h_1\to\gamma\gamma)
\sim
\sigma(pp\to Va)\,{\rm BR}(a\to\gamma\gamma),
\end{align}
In Fig.~\ref{fig:disvar}, we show the reconstructed diphoton invariant-mass
distribution for both signal hypotheses, together with the SM backgrounds, for
the representative mass point $m_a=m_{h_1}=80~{\rm GeV}$. 
This plot is  subject to pre-selection Eq.~\eqref{eq:gencuts} and $\Delta_\phi(\gamma_1,\gamma_2)\leq 1.5$ cuts. The mass window is defined from the signal mass resolution, taken as the smallest interval containing $68.3\%$ of the reconstructed
$m_{\gamma\gamma}$ distribution.  The same window is then applied to the
background samples. This mass window contains $284$ signal events for each signal hypothesis and about $780$ total SM background events, with
the dominant contribution coming from $W\gamma j$.

\begin{figure}[htp!]
        \centering
\includegraphics[width=0.48\linewidth]{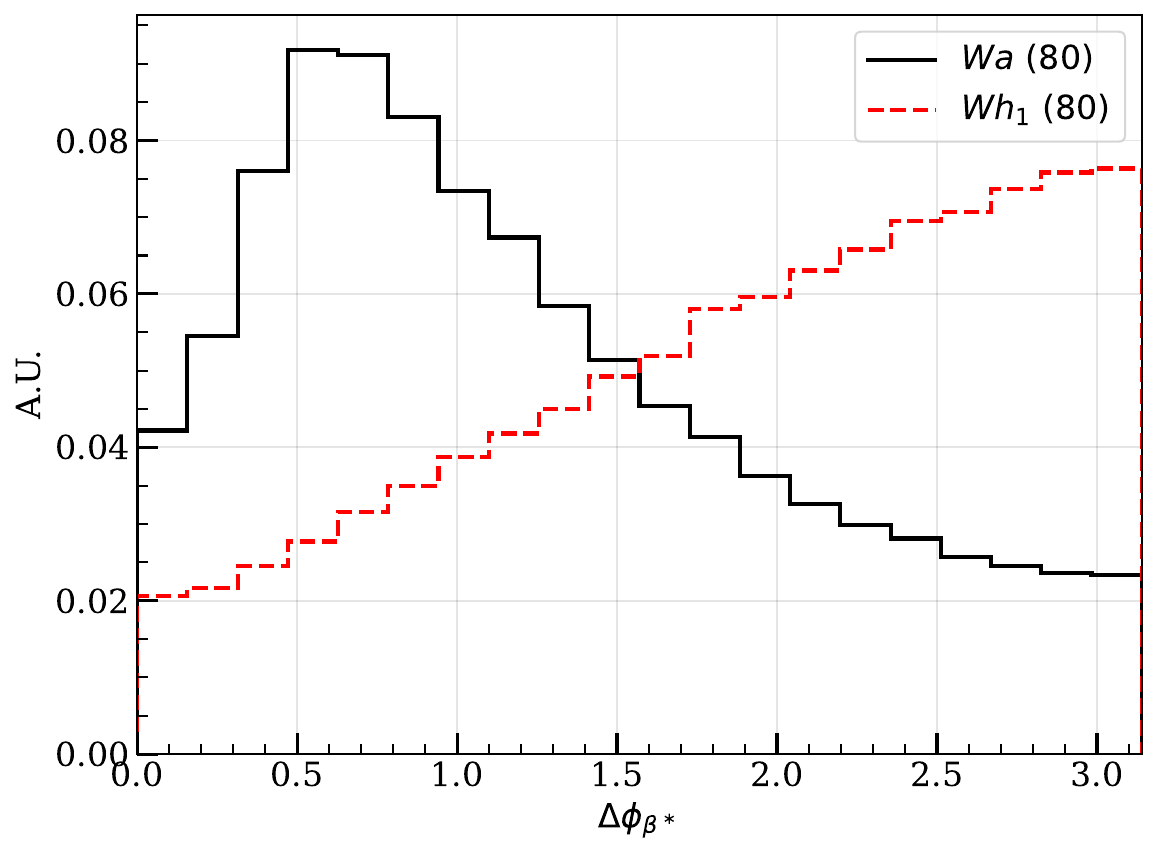}       
\includegraphics[width=0.48\linewidth]{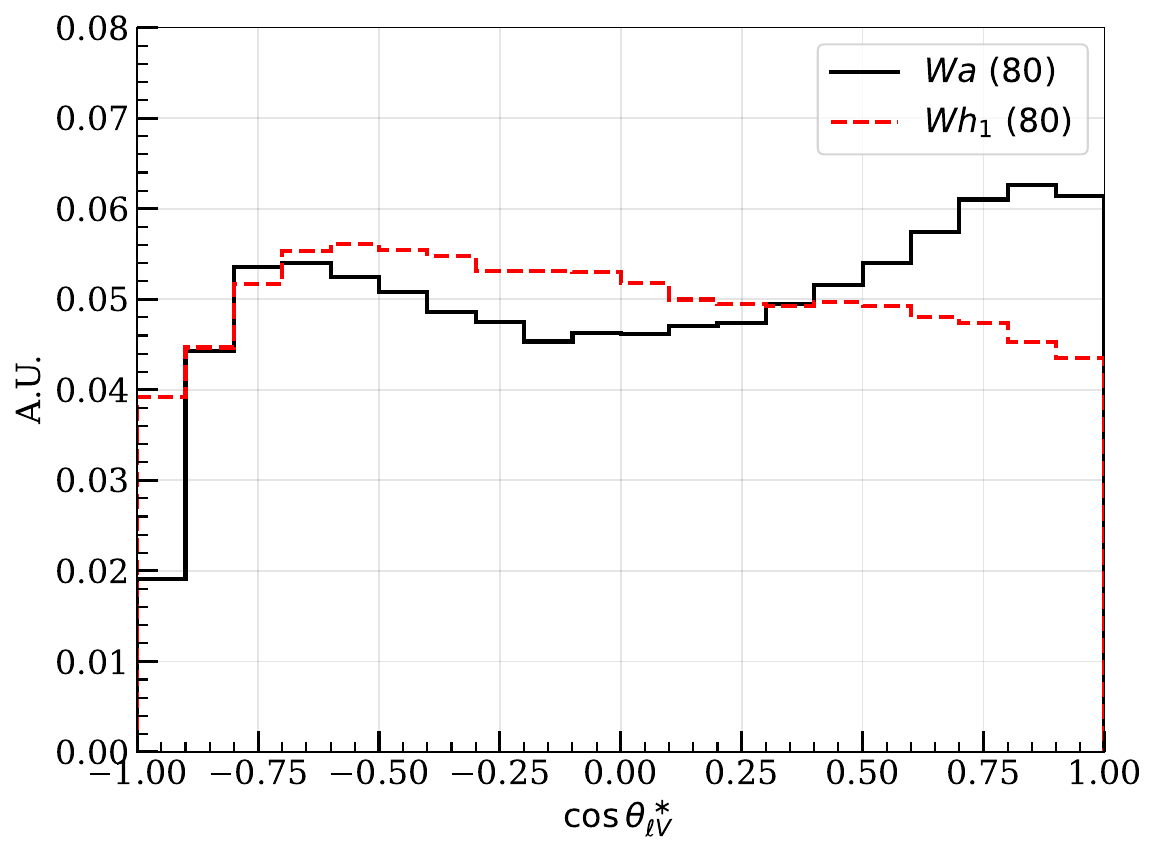}   
 \caption{Normalized
$\Delta \phi_\beta^\ast$ (left) $\cos\theta_{\ell V}^\ast$ (right) distributions for the $Wa$ and $Wh_1$ hypotheses with $m_a=m_{h_1}=80~{\rm GeV}$ for the events which lie within the mass
window shown in Fig.~\ref{fig:disvar}.}
\label{fig:disvar1}
\end{figure}

\begin{figure}[htp!]
\centering
\includegraphics[width=\linewidth]{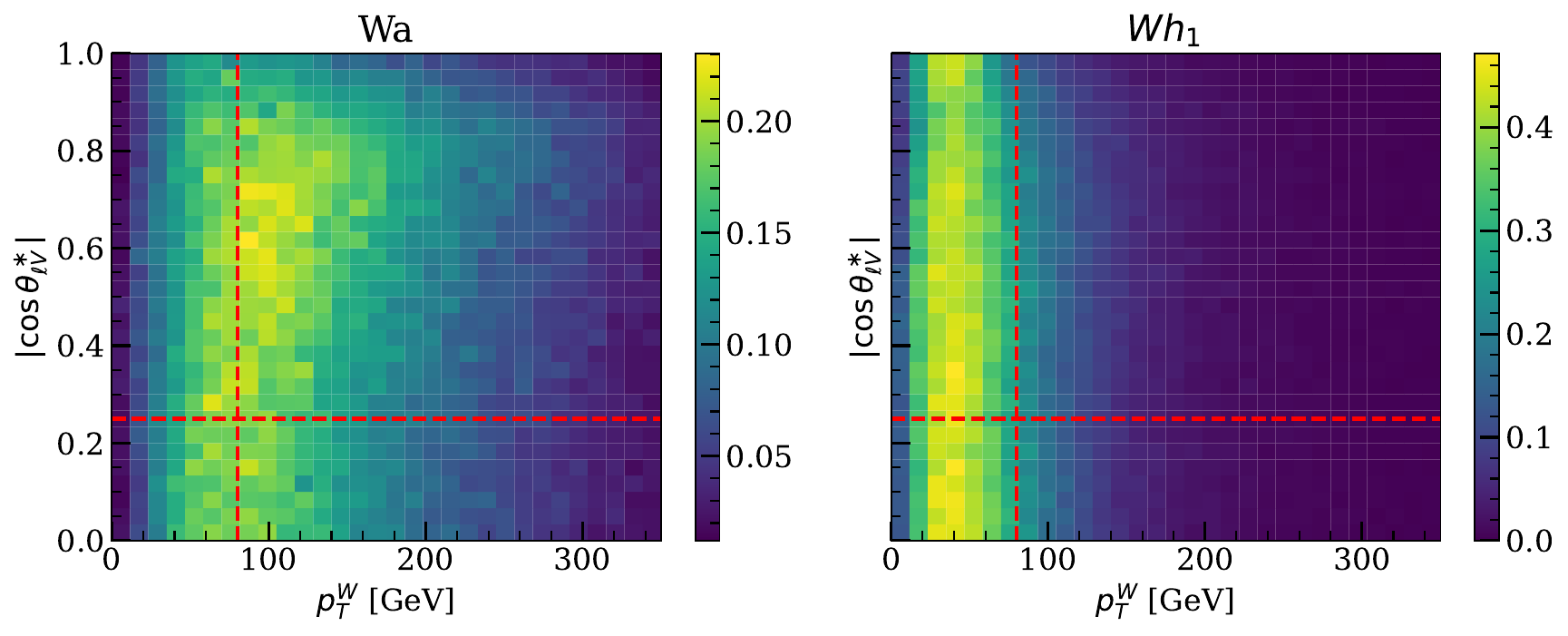} 
\includegraphics[width=\linewidth]{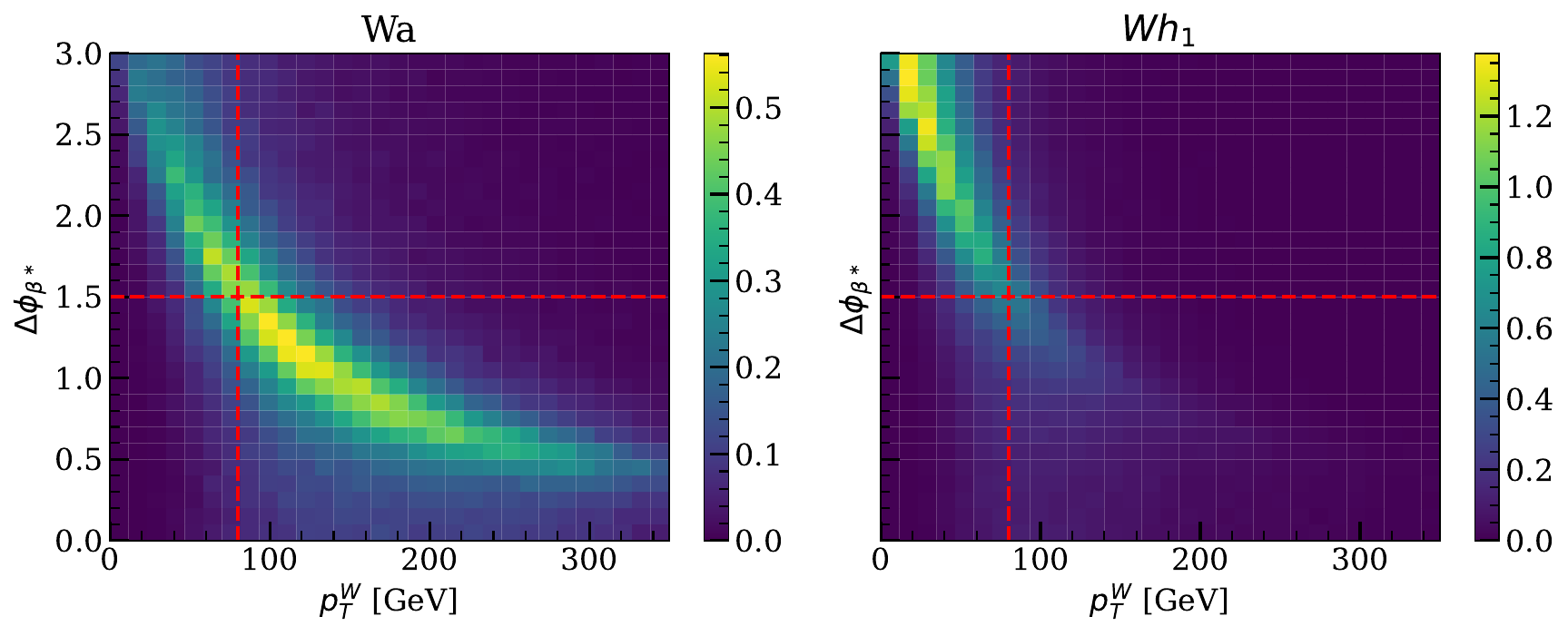} 
	\caption{Normalized two-dimensional distributions in the
$p_T^W$--$|\cos\theta_{\ell V}^\ast|$ (top row) and $p_T^W$--$\Delta\phi_{\beta^\ast}$ (bottom row) planes for the $Wa$ (left) and $Wh_1$ (right) hypotheses with $m_a=m_{h_1}=80~{\rm GeV}$.}
	\label{fig:2d_pt_costheta}
\end{figure}

We now turn to the polarization- and recoil-sensitive observables, in particular the angular variables $\Delta \phi_{\beta^\ast}$, $\cos\theta_{\ell V}^\ast$, etc., and the transverse momentum of the associated gauge boson, $p_T^W$. Fig.~\ref {fig:disvar1} shows $\Delta \phi_{\beta^\ast}$(left), as defined in Eq.~\eqref{eq:dphi_betastar}, and $\cos\theta_{\ell V}^\ast$ (right) distributions for the $Wa$ and $Wh_1$ samples for events restricted within the diphoton invariant-mass window shown in Fig.~\ref{fig:disvar}. The two models show a very clearly distinctive pattern, in particular $\Delta \phi_\beta^\ast > 1.5$ by a factor of 2--4 and in two regions of 
$-0.5 < \cos\theta^\ast_{\ell V} < 0.4$ , $\cos\theta^\ast_{\ell V}>$0.4
by factor $\sim$ 1.5.
This difference reflects the distinct polarization compositions of the associated $W$ boson in the two production mechanisms.
As discussed above, $Wa$ production is dominated by transverse $W$ polarization, with a strongly suppressed longitudinal component, whereas $Wh_1$ production contains a
substantially larger longitudinal contribution.
In addition, the recoil momentum of the $W$ boson provides another useful and complementary feature for discriminating between these two models.
With this observation, the distributions of events 
are presented 
that are within the mass window are presented in two-dimensional (2D) $p_T^W - |\cos\theta^\ast_{\ell V}|$, $p_T^W - \Delta\phi_{\beta^\ast}$ planes for $m_a=80$~GeV, shown in Fig.~\ref{fig:2d_pt_costheta}. This representation combines the recoil and polarization-sensitive information and reveals a clear difference in the regions populated by the $Wa$ and $Wh_1$ signals. Appropriate selection requirements, such as $p_T^W>80~{\rm GeV}$, $|\cos\theta_{\ell V}^\ast|>0.25$, $\Delta\phi_{\beta^\ast} < 1.5$ along with other variables like , $0<LP<1$ populate the Wa signal events and hence 
can therefore effectively separate the two signal models. We find that, out of 284 signal events, 169 and 111 events remain for the $Wa$ and $Wh_1$ hypotheses, respectively, while 158 background events survive.
Table~\ref{tab:cutflow_model} summarizes the corresponding expected
significances $\mathcal{Z}$ for the mass-matched $m_a=m_{h_1}=80~{\rm GeV}$ benchmark.
Treating each signal hypothesis in turn as the signal, with the SM contribution as the background, we
obtain $\mathcal{Z}=6.7$ for $Wa$ and $\mathcal{Z}=4.7$ for $Wh_1$. The resulting significances indicate that the combined recoil- and
polarization-sensitive selections provide considerable discrimination between
the two signal hypotheses, even when their masses and overall signal rates are
matched.
\begin{table}[htb!]
\centering
\begin{tabular}{|l|cccc|}
\hline
 \textbf{$C_{WW}$} & $m_a=m_{h_1}$ & $Wa$/$Wh_1$ & SM Bg 
& $\mathcal{Z} = Wa/Wh_1$\\
\hline
0.02 $\mathrm{TeV}^{-1}$ & 80 & 169/111 & 158 & 6.7/4.7 \\
\hline   
\end{tabular}
\caption{Expected yields and corresponding significances at the HL-LHC for the $Wa$
and $Wh_1$ hypotheses with $m_a=m_{h_1}=80~{\rm GeV}$ and
${\cal L}=3~{\rm ab}^{-1}$. The two signal hypotheses are normalized to equal yields after the baseline selections and $\Delta\phi(\gamma_1,\gamma_2)<1.5$.
The quoted yields are obtained after additionally imposing
$p_T^W>80~{\rm GeV}$, $|\cos\theta_{\ell V}^\ast|>0.25$,
$\Delta\phi_{\beta^\ast} < 1.5$, and $0<LP<1$ within the diphoton mass
window $[71.6,88.4]~{\rm GeV}$. The significance is evaluated for each
hypothesis in turn, treating the SM contribution as background and assuming a $10\%$ background systematic
uncertainty.}
\label{tab:cutflow_model}
\end{table}

\begin{figure}[htp!]
  \centering
  \includegraphics[width=0.6\linewidth]{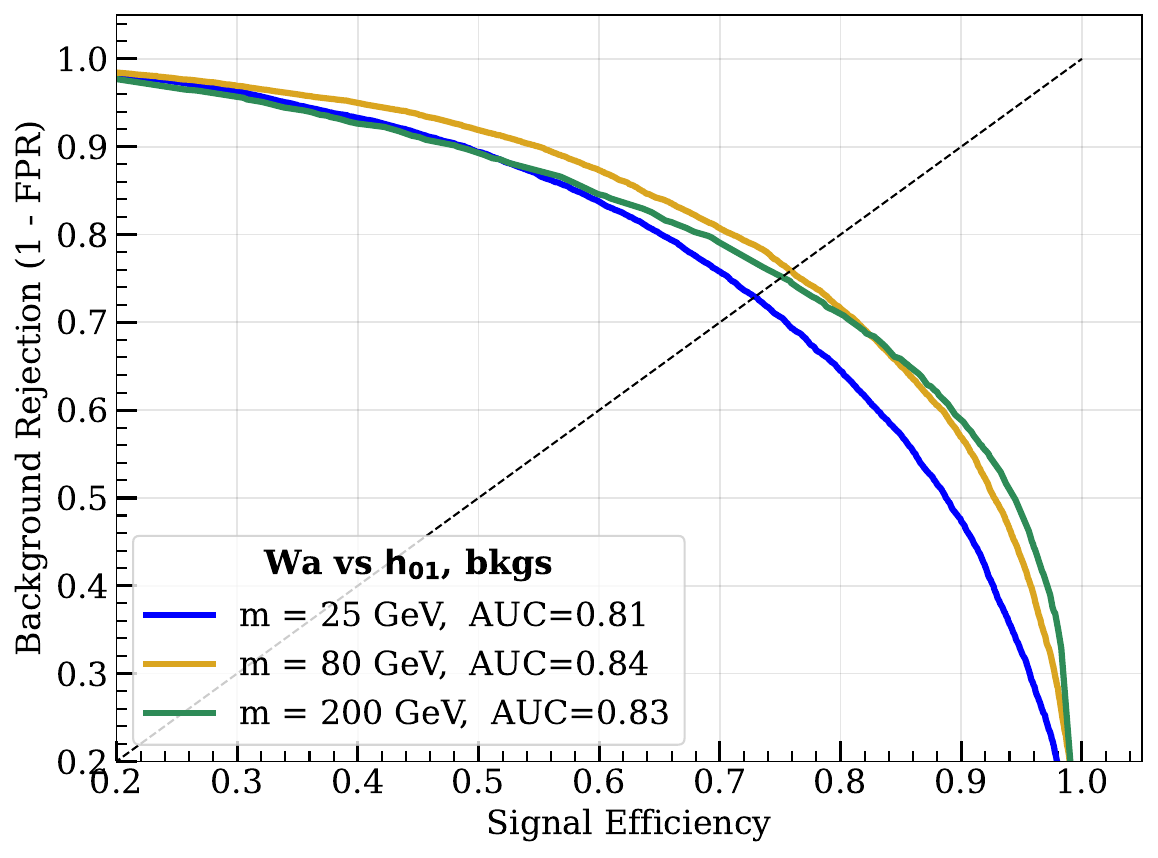}
    \caption{Receiver operating characteristic (ROC) curves for discriminating the $Wa$
hypothesis from the mass-matched $Wh_1$ hypothesis together with the SM
backgrounds, for $m=25$, $80$, and $200~{\rm GeV}$.}
    \label{fig:rocs_toy}
\end{figure}

To further improve the discrimination between the two signal hypotheses, we perform an XGBoost-based multivariate analysis using the events within the di-photon invariant-mass window. 
Using a training strategy similar to that described
in the previous section, we construct an XGBoost classifier to discriminate between the $Wa$ and $Wh_1$ hypotheses.
Fig.~\ref{fig:rocs_toy} shows the corresponding ROC curve for an XGBoost classifier trained to separate the ALP hypothesis ($Wa$) from the mass-matched $Wh_1$ sample together with the SM backgrounds for three representative choices of ALP masses $m_a=20, 80, 200$~GeV. The obtained AUC $\sim$ 0.80 demonstrates that sizable discrimination remains even
when the two signal hypotheses produce the same reconstructed di-photon mass peak. 
Evidently, information about the different
Lorentz structures of the underlying interactions is encoded in the event-level kinematics and can be exploited to probe the nature of an observed spin--0 resonance.

\section{Summary}
\label{sec:conclusions}

The associated production of an axion-like
particle with an electroweak gauge boson at the HL-LHC is studied
with center-of-mass energy  
$\sqrt{s}=14~{\rm TeV}$,  and target luminosity
of $\mathcal L=3~{\rm ab}^{-1}$. We focus
on the
$W^\pm a$ and $Za$ channels with the subsequent decay $a\to\gamma\gamma$. Results are presented 
for a selected electroweak ALP
benchmarks in which the gluon and fermion couplings with ALPs 
are set to zero, to ensure that
the production and decay are governed by the ALP couplings to electroweak
gauge bosons only.  The ALP mass is varied in the range
$20~{\rm GeV}\le m_a\le 200~{\rm GeV}$, where the diphoton final state can be
treated predominantly in the resolved-photon regime.

The main observation underlying our analysis is that the tensor structure of
the ALP interaction leaves characteristic signatures in the kinematics of the
associated gauge boson.  The momentum-dependent
$aV_{\mu\nu}\widetilde V^{\mu\nu}$ interaction leads to a harder recoil
spectrum and a dominantly transverse polarization pattern for the associated
gauge boson at the on-shell Born level.  
In contrast, it is very qualitatively different
from a Higgs-like non-derivative $HV_\mu V^\mu$ interaction, which shows a larger
longitudinal component, in particular in the high transverse momentum regime. These characteristic features are used to construct observables based on the diphoton recoil, the transverse momentum of the associated gauge boson, and polarization-sensitive angular variables using lepton momentum.

We first performed a cut-based analysis for both the 
($W^\pm$a) and
(Za) channels involving charged and neutral currents, respectively. 
The dominant backgrounds were taken to be continuum
$V\gamma\gamma$ production, reducible $V\gamma j$ production with a jet
misidentified as a photon, and associated SM Higgs production(VH) 
with
$H\to\gamma\gamma$.  The diphoton recoil requirement provides the strongest
single handle for suppressing the continuum backgrounds 
while retaining a
sizable fraction of the signal.  The angular observables, such as
$\Delta\phi(\gamma_1,\gamma_2)$, $\cos\theta^\ast_{\ell V(\ell\ell)}$  
further improve the
signal-to-background separation and, more importantly, retain information about
the production dynamics.  For the benchmark choices considered, the
$W^\pm a$ channel gives the stronger discovery reach because of its larger
production rate, while the $Za$ channel provides a complementary probe of the
neutral electroweak ALP coupling combinations.

Further to improve the sensitivity, the analysis is performed 
using a multivariate classifier based on
\texttt{XGBoost}.  Two training strategies were considered.  In the first, the
diphoton invariant mass was excluded from the input variables in order to test
how much discrimination can be obtained from non-resonant kinematic, recoil and
polarization information alone.  In the second case, $m_{\gamma\gamma}$ was included
as an additional input feature, corresponding to a mass-assisted analysis at a
fixed ALP mass hypothesis.  In the first case, the classifier shows good signal--background separation
even without using the diphoton mass peak, with AUC values typically close to
or above $0.9$ for the benchmark masses studied.  Including
$m_{\gamma\gamma}$ further improves the performance, with AUC values close to
unity in several cases.  
This demonstrates that the conventional diphoton resonance information can be significantly complemented by observables that are directly sensitive to the underlying production dynamics.

Using the expected signal and background yields from the multivariate analysis,
the reach of sensitivities on the electroweak ALP
couplings 
are obtained.
For the charged-current channel, the reach can be directly
interpreted as a sensitivity to the $aWW$ interaction, or equivalently to
$C_{WW}/f_a$.  For the neutral-current
channel, the interpretation is necessarily benchmark-dependent because
$Za$ production receives coherent contributions from both the $aZZ$ and
$aZ\gamma$ interactions.  We therefore considered a fixed direction in the
$(C_{WW},C_{BB})$ plane.  For representative mass points, the projected
HL-LHC sensitivity reaches the level of $10^{-6}$--$10^{-5}~{\rm GeV}^{-1}$,
depending on the channel, mass and coupling direction.

Finally, we explored whether the same final state can be used not only to
discover a resonance, but also to characterize its underlying dynamics.
For this purpose, the framework based on the NMSSM model 
is considered as a benchmark model, which provides multiple 
Higgs boson states, of which one or two states can be ALP -like depending on the model.
To
this end, we compared the ALP signal with a mass-matched Higgs-like spin--0
look-alike whose visible $V\gamma\gamma$ rate was normalized to match the ALP
benchmark.
This removes the trivial discrimination from the total rate and
focuses the comparison on shape information.  The ALP and Higgs-like
hypotheses populate the recoil and helicity-angle observables differently,
reflecting the different Lorentz structures of their production vertices.  A
dedicated multivariate analysis retains nontrivial discrimination power, with
AUC values around $0.8$ for the representative benchmarks considered.  This
shows that recoil and polarization information can provide useful handles for
interpreting the nature of a possible diphoton resonance.

Overall, our results show that for the associated production with electroweak gauge
bosons, the accompanying gauge boson is not merely a trigger or background-suppression handle; its
polarization and recoil properties encode information about the tensor
structure of the spin--0 interaction. These observables can improve the
sensitivity to ALP signals and can help distinguish an ALP-like interaction
from Higgs-like spin--0 look-alike scenarios. 
However, it is worth noting that the robust features of polarization-sensitive observables, such as $\cos\theta^\ast_{\ell V}$ and other variables defined in the helicity frame, can be significantly diluted once experimental selection requirements are imposed. Consequently, the full discrimination power of these observables cannot be fully utilized at the reconstructed event level. This motivates the use of low-level information for particle reconstruction, such as energy deposits in the calorimeter and tracking information, as inputs to machine-learning(ML) algorithms. May be this approach based on ML can potentially recover additional information that is lost at the reconstruction level and substantially enhance the discriminating power of the analysis [A. De, M. Guchait, A. Roy, in preparation].

\section*{Acknowledgements}
The work of A.R. is supported by an Australian Research Council Discovery Project, ARC-DP230101142. A.R. thanks Basudeb Dasgupta and the Department of Theoretical Physics, Tata Institute of Fundamental Research, for hospitality, local support, and accommodation, and Ranjan Laha and the Centre for High Energy Physics, Indian Institute of Science, Bengaluru, for hospitality during visits where part of this work was carried out and discussed. AR also acknowledges support from ICCUB, where he joined during the final stages of writing this manuscript, towards the completion of this work. MG gratefully acknowledges the Theory Division of CERN, where this work was initiated. MG also thanks the Korea Institute for Advanced Study (KIAS), Seoul, where a substantial part of this work was completed.

\section{Appendix}
\subsection*{$W$ momentum reconstruction}
\label{sec:App_W_momenta}
The reconstruction of the neutrino momentum in $W \to \ell \nu$ decays is crucial to this analysis as it directly impacts the angular observables such as $\cos\theta^*$, especially in the forward and backward regions ($|\cos\theta^*| \to 1$) which are of primary interest for polarization and CP studies.  Since only the charged lepton is directly measured, the neutrino four-momentum $p_\nu=(p_{\nu_x},p_{\nu_y},p_{\nu_z},E_\nu)$ must be inferred from the missing transverse momentum
($\vec{p}_{T}^{\,\rm miss}$) together with the on-shell $W$ mass constraint \cite{Grossi:2020orx}. The steps of reconstruction are as follows
\begin{enumerate}
    \item The measured $\met$ provides the transverse components of the neutrino momentum, which are allowed to vary within a small window to account for detector resolution, as $p_{\nu_x} = p_{\nu_x} + \delta_x,~p_{\nu_y} = p_{\nu_y} + \delta_y$ where $\delta_x,\delta_y \in [-3,+3]$ GeV with steps of 1 GeV.
    \item For each $(p_{\nu_x}, p_{\nu_y})$, solve the quadratic equation from $(p_\ell + p_\nu)^2 = M_W^2$ as,
    \begin{equation}
        ap_{\nu_z}^2 + b p_{\nu_z} + c = 0 
    \end{equation}
    where $a = E_\ell^2 - p_{\ell_z}^2$, $b = -2\mu p_{\ell_z}$, $c = \mu^2 - E_\ell^2|\vec{p}_T^\nu|^2$, and $\mu = \frac{M_W^2 - m_\ell^2}{2} + \vec{p}_T^\ell\cdot\vec{p}_T^\nu$. Both real solutions for $p_{\nu_z}$ are retained for further evaluation.
    
    \item For each candidate neutrino momentum ($p_{\nu_z}^{1,2}$) we construct the reconstructed $W$ boson,$p_W=p_\ell+p_\nu$, and evaluate the likelihood through,
        \begin{equation}
         \chi^2 = \chi^2_{\text{MET}} + \chi^2_W  + \chi^2_{p_{\nu_z}}
                =\underbrace{\frac{\delta_x^2 + \delta_y^2}{\sigma_{\text {MET}}^2}}_{\text {MET resolution }}+\underbrace{\frac{\left(m_{\ell v}-M_W\right)^2}{\sigma_W^2}}_{\text {W mass }}+\underbrace{\frac{p_{\nu_z}^2}{\sigma_{p_{\nu_z}}^2}}_{p_{\nu_z}~\text{regularization}}
        \end{equation}
    The First term penalizes deviations from the measured MET, ensuring consistency with detector-level information, $\sigma_{\text{MET}} =3$ GeV. The second term enforces consistency with the $W$ boson mass, where $\sigma_W = 2.1$ GeV is considered, reflecting detector resolution and finite width. The third term suppresses unphysically large $p_{\nu_z}$ solutions in the regions with large kinematic ambiguity, $\sigma_{p_{\nu_z}} = 100$ GeV. 
    
    \item The neutrino candidate minimizing $\chi^2$ is selected. If two solutions have nearly identical $\chi^2$, the one with smaller $p_{\nu_z}$ is chosen, and if no real solutions exist in the grid, use the collinear approximation $p_{\nu_z} = -\frac{b}{2a}$, which corresponds to setting the discriminant to zero.
\end{enumerate}

\bibliographystyle{JHEP}
\bibliography{References}
\end{document}